\documentclass[12pt,a4paper]{article}

\usepackage[T1]{fontenc} 				

\usepackage{amssymb,amsmath,amscd}
\usepackage[english]{babel}
\usepackage{titlesec}
\usepackage{slashed}
\usepackage{hyperref}
\usepackage{graphicx,xcolor}
\usepackage{enumitem}
\usepackage{multirow}
\usepackage{placeins}
\usepackage[framemethod=tikz]{mdframed}
\usepackage[absolute]{textpos}
\usepackage[margin=1.5em,labelfont=bf]{caption}

\hypersetup{
	colorlinks=true,
	linkcolor=dark-blue,
	citecolor=dark-red,
	urlcolor=dark-green,
	linktoc=all
}

\graphicspath{{./figures/}}

\usepackage{geometry} 					
\numberwithin{equation}{section} 

\titleformat{\section}[block]{\Large\bfseries}{\thesection}{1em}{} 
\titleformat{\subsection}[block]{\bfseries}{\thesubsection}{1em}{} 
\titlespacing*{\section}{0pt}{1em}{1em}
\titlespacing*{\subsection}{0pt}{0.75em}{0.75em}

\graphicspath{{./figures/}}

\newcommand\Tstrut{\rule{0pt}{2.6ex}}         
\newcommand\Bstrut{\rule[-0.9ex]{0pt}{0pt}}   

\newmdenv[innerlinewidth=0.5pt, roundcorner=1pt,linecolor=dark-blue,innerleftmargin=6pt,
innerrightmargin=6pt,innertopmargin=6pt,innerbottommargin=6pt]{mybox}

\definecolor{dark-gray}{gray}{0.20}
\definecolor{gray}{gray}{0.30}
\definecolor{light-gray}{gray}{0.80}
\definecolor{dark-red}{rgb}{0.7,0,0}
\definecolor{dark-green}{rgb}{0.1,0.4,0}
\definecolor{dark-blue}{rgb}{0.3,0.3,0.7}
\definecolor{light-blue}{rgb}{0.8,0.8,1}
\definecolor{cardinal}{rgb}{0.6,0,0}
\definecolor{darkgreen}{rgb}{0,0.5,0}
\definecolor{golden}{rgb}{0.92, 0.7, 0}
\definecolor{midnight}{rgb}{0, 0, 0.5}
\definecolor{darkblue}{rgb}{0.2, 0, 0.8}
\definecolor{forestgreen}{rgb}{0.13, 0.55, 0.13}

\newcommand\bN{\mathbf{N}}

\newcommand\bC{\mathbf{C}}
\newcommand\bP{\mathbf{P}}
\newcommand\bR{\mathbf{R}}
\newcommand\bZ{\mathbf{Z}}

\newcommand\cM{\mathcal{M}}
\newcommand\cN{\mathcal{N}}

\newcommand{\dd}{\mathrm{d}}
\newcommand{\e}{\mathrm{e}}

\newcommand{\dvol}{{\rm vol}}

\renewcommand{\gcd}{\mathrm{gcd}}

\newcommand\AdS{\mathrm{AdS}}

\newcommand\Hom{\mathrm{Hom}}

\newcommand{\f}[2]{\frac{#1}{#2}}

\newcommand{\nn}{\nonumber}
\newcommand{\ds}{{\rm d}s}

\newcommand {\be} {\begin {equation}}
\newcommand {\ee} {\end {equation}}
\newcommand {\bes} {\begin {equation*}}
\newcommand {\ees} {\end {equation*}}

\newcommand\SO{\mathrm{SO}}

\newcommand\UU{\mathrm{U}}
\newcommand\SU{\mathrm{SU}}

\newcommand\Spin{\mathrm{Spin}}
\newcommand\Cliff{\mathrm{Cliff}}

\newcommand\so{\mathfrak{so}}
\newcommand\su{\mathfrak{su}}

\newcommand\osp{\mathfrak{osp}}

\title{\vspace{10mm}\fontsize{22pt}{22pt}\selectfont\textbf{Spin Structures and AdS$_4$ Holography}\vspace{3mm}}

\author{\large Nikolay Bobev$^{\dagger}$ and  Pieter Bomans$^{\odot\bullet\circ}$\rule{0pt}{1.5em}
}

\date{}

\begin{document}  

	\maketitle
	\thispagestyle{empty}
	
	\vspace{0em}
	
	\begin{center}
		\begin{tabular}{rl}
			$^{\dagger}$ & \hspace{-3mm}Instituut voor Theoretische Fysica, KU Leuven\\
			& \hspace{-3mm}Celestijnenlaan 200D, BE-3001 Leuven, Belgium\\[2mm]
			$^{\odot}$ & \hspace{-3mm}Joseph Henry Laboratories, Princeton University\\
			& \hspace{-3mm}Princeton, NJ 08544, USA\\[2mm]
			$^{\bullet}$ & \hspace{-3mm}Dipartimento di Fisica e Astronomia, Universit\`a di Padova\\
			& \hspace{-3mm}via Marzolo 8, 35131 Padova, Italy\\[2mm]
			$^{\circ}$ & \hspace{-3mm}INFN, Sezione di Padova\\
			& \hspace{-3mm}via Marzolo 8, 35131 Padova, Italy\\[10mm]
		\end{tabular}
		
		\texttt{\textcolor{dark-green}{\href{mailto:nikolay.bobev@kuleuven.be}{nikolay.bobev@kuleuven.be} \,\,\, \href{mailto:pieter.bomans@gmail.com}{pieter.bomans@gmail.com}  }}	
	\end{center}
	
	\vspace{2em}
	
	\begin{abstract}
		\noindent We study the importance of spin structures as defining data for 11d supergravity backgrounds of the form $\AdS_4\times S^7/\bZ_k$ with a free orbifold action. For a generic choice of the orbifold action, there is only one spin structure that preserves invariant Killing spinors. There are, however, orbifold actions that allow for more than one spin structure with preserved Killing spinors. In these cases the two different spin structures should lead to distinct holographically dual 3d SCFTs with different amounts of supersymmetry. We illustrate this phenomenon by studying the KK spectrum of 11d supergravity on $\AdS_4\times S^7/\bZ_4$. For one choice of spin structure this background is dual to the 3d $\mathcal{N}=6$ ABJM theory at level $k=4$. For the alternative choice of spin structure our results suggest that the holographic dual is a different 3d $\cN=2$ SCFT and we discuss some of its properties. 
	\end{abstract}
	
	\clearpage
	\setcounter{page}{1}
	
	{
		\hypersetup{linkcolor=black}
		\tableofcontents
	}

\section{Introduction}
\label{sec:intro}

Spin structures on (pseudo-)Riemannian manifolds have a wide range of applications in theoretical physics, most importantly in the study of any QFT containing spinors. In particular, the spin structure determines the periodicity of spinors around non-trivial loops in the geometry. For this reason, a careful choice of the spin bundle constitutes a fundamental part of the defining data of any QFT. For instance, in 2d CFTs the choice of spin structure distinguishes between the Neveu-Schwarz and Ramond sectors, while in string theory, summing over the spin structures amounts to performing the GSO projection \cite{Seiberg:1986by}. Similarly, in three-dimensional spin TQFTs, the partition function depends on the choice of spin structure, hence a careful treatment of the various spin structures is essential in the study of topological phases of matter, see for example \cite{Dijkgraaf:1989pz,Kapustin:2017jrc,Karch:2019lnn,Okuda:2020fyl}. 

Spin structures play an important role also in supergravity. For example, as discussed in \cite{Witten:1981gj}, the choice of spin structure for the KK vacuum $\bR^{1,3}\times S^1$ determines its non-perturbative instability if there are elementary fermions in the gravitational theory. This suggests that in the context of AdS/CFT the choice of spin structure can have significant implications for the dual field theory. Our goal here is to discuss a particular situation in supergravity where the choice of spin structure has important implications in the context of AdS$_4$/CFT$_3$ holography. 

The setup we consider is the well-known AdS$_4\times S^7$ solution of 11d supergravity, supported by $N$ units of flux. This background can be obtained as the near horizon limit of the supergravity solution describing $N$ coincident M2-branes in $\mathbf{C}^4$. This solution is maximally supersymmetric and is holographically dual to the ABJM theory, which is a 3d $\mathcal{N}=8$ SCFT describing a $\UU(N)\times \UU(N)$ gauge theory with Chern-Simons level $k=1$~\cite{Aharony:2008ug}. This setup admits a simple generalization in which the M2-branes probe the space $\mathbf{C}^4/\bZ_k$ where the orbifold action acts with an equal phase $\e^{2\pi i / k}$ on all four complex coordinates of $\mathbf{C}^4$, where $k$ is a positive integer. The near horizon limit of the backreacted solution in this case is of the form AdS$_4\times S^7/\bZ_k$ and since the orbifold action is free the supergravity solution is smooth. As discussed in \cite{Figueroa-OFarrill:2005vxy}, and summarized below in Section~\ref{sec:spin} for odd values of $k>1$ there is only one choice of spin structure on $S^7/\bZ_k$. The spin structure is compatible with the 6 out of the 8 Killing spinors on $S^7$ and we therefore conclude that the AdS$_4\times S^7/\bZ_k$ 11d supergravity background is dual to a 3d SCFT with $\mathcal{N}=6$. Indeed, this is the well known $\UU(N)\times \UU(N)$ ABJM theory at level $k$ which is known to preserve $\mathcal{N}=6$ supersymmetry for odd $k>1$. The situation is more interesting for even values of $k$. In this case there are two distinct spin structures on $S^7/\bZ_k$ which we refer to as \textit{periodic} and \textit{antiperiodic}, depending on the periodicity of the fermions along the non-trivial loop in the geometry. For all even values of $k>4$ the periodic spin structure preserves 6 out of the 8 Killing spinors on $S^7$ and the dual SCFT is the $\mathcal{N}=6$ $\UU(N)\times \UU(N)$ ABJM theory at level $k$. For even $k>4$ the antiperiodic spin structure does not preserve any Killing spinors and therefore these AdS$_4\times S^7/\bZ_{k}$ backgrounds should be dual to non-supersymmetric CFTs. In the absence of supersymmetry, and in view of the conjecture in \cite{Ooguri:2016pdq}, it is unclear whether this supergravity background is stable. We therefore do not discuss this situation further. For $k=2$ or $k=4$ both spin structures preserve some number of the Killing spinors on $S^7$. When $k=2$, we have $S^7/\bZ_2= \bR\bP^7$ for which both spin structures preserve the maximal number of Killing spinors \cite{FRANC1987277}. In this case we expect that the AdS$_4\times S^7/\bZ_2$ supergravity solution for both choices of spin structure is dual to the $\UU(N)\times \UU(N)$ ABJM theory at level $k=2$ which is known to exhibit supersymmetry enhancement to $\mathcal{N}=8$ due to the presence of light monopole operators \cite{Bashkirov:2010kz}. The case $k=4$ is more subtle. On $S^7/\bZ_4$ the periodic spin structure preserves 6 invariant Killing spinors while the antiperiodic one preserves 2. Put differently, the lens space $S^7/\bZ_4$ preserves 8 invariant Killing spinors, but 2 of them live in a different spin bundle from the other 6. This discussion therefore suggests that the AdS$_4\times S^7/\bZ_4$ supergravity solution is dual to the $\mathcal{N}=6$ $\UU(N)\times \UU(N)$ ABJM theory at CS level $k=4$ when we choose the periodic spin structure and is dual to some other $\cN=2$ SCFT when we choose the antiperiodic one.

The fact that there are two distinct SCFTs dual to the same AdS$_4\times S^7/\bZ_4$ solution of 11d supergravity appears somewhat exotic and deserves further scrutiny. To understand the properties of these two SCFTs better we study the spectrum of masses for the KK modes on AdS$_4\times S^7/\bZ_4$ for both choices of spin structure. The holographic dictionary dictates that these modes should be dual to single trace operators in the dual SCFT which should organize in superconformal multiplets. To find the explicit KK spectrum we utilize the well known result for the spectrum of 11d supergravity excitations around AdS$_4\times S^7$, see \cite{Duff:1986hr,Biran:1983iy}, and then use the $\bZ_4$ action to project onto the invariant modes. The bosonic modes that survive this projection are the same for both choices of spin structures. The fermionic modes are however different. For the positive spin structure we show that all the bosonic and fermionic modes organize into $\mathcal{N}=6$ superconformal multiplets and lead to the expected spectrum of single trace operators for the $\mathcal{N}=6$ $\UU(N)\times \UU(N)$ ABJM theory at CS level $k=4$. For the antiperiodic spin structure we find that all KK modes organize into $\mathcal{N}=2$ superconformal multiplets and furthermore find that the multiplets fall into representations of an $\SU(4)$ flavor symmetry. This is compatible with the fact that the $\bZ_4$ orbifold action preserves an $\SU(4)$ subgroup of $\SO(8)$ which furthermore leaves the two Killing spinors of the antiperiodic spin structure invariant. Notably, we find that the spectrum of operators corresponding to the KK modes in this $\mathcal{N}=2$ SCFT contains both short and long multiplets. This analysis suggests that the new $\cN=2$ SCFT is very closely related to the $\mathcal{N}=6$ $\UU(N)\times \UU(N)$ ABJM theory at level $k$, at least when it comes to the spectrum of operators dual to the supergravity KK modes. The two theories share identical spectrum of bosonic operators but they have different fermionic operators which is ultimately responsible for the different amount of supersymmetry preserved by the two models.

There are many possible orbifolds of AdS$_4\times S^7$ and it is natural to wonder if the dichotomy of spin structures we discussed above is present also in other examples. Indeed, as discussed in \cite{Figueroa-OFarrill:2004lpm,Figueroa-OFarrill:2005vxy,Gadhia:2007lxa,deMedeiros:2009pp} there are numerous other situations in which orbifolds of $S^7$ allow for two spin structures, each of which preserves some of the Killing spinors on $S^7$. We illustrate the rich structure of these orbifolds by studying some examples of AdS$_4\times S^7/\bZ_k$ smooth orbifolds, i.e. lens spaces, where the $\bZ_k$ action differs from the one above and preserves only a $\SU(2)\times\SU(2)\times \UU(1)^2$ subgroup of $\SO(8)$. For specific choices of the orbifold action one again finds that there are two possible spin structures. One of them leads to an AdS$_4\times S^7/\bZ_k$ supergravity solution dual to a 3d $\mathcal{N}=4$ SCFT while the other should be dual to a 3d $\mathcal{N}=2$ theory. We describe some of these examples in detail and show that the KK spectrum of 11d supergravity is indeed compatible with the existence of such pairs of SCFTs with identical bosonic spectra but different amount of supersymmetry.

In the remainder of this paper we study the $S^7/\bZ_4$ example in more detail and also discuss some generalizations to other lens spaces. In Section~\ref{sec:spin} we start with a brief introduction to spin structures and in particular describe the possible spin structures of $S_7/\bZ_k$. We proceed in Section~\ref{sec:KK} with a discussion of the KK spectrum of these 11d supergravity backgrounds and organize it in superconformal multiplets for both choices of spin structure that preserve some amount of supersymmetry. In Section~\ref{sec:discussion} we discuss some aspects of the dual SCFT and several interesting questions for future work. In the three appendices we collect our conventions together with some additional group theory and superconformal representation theory details and examples.

\section{Spin structures and supersymmetry}
\label{sec:spin}

An orientable $n$-dimensional manifold $\cM$ is spin whenever its second Stiefel-Whitney class $w_2\in H^2(\cM;\bZ_2)$ vanishes. In this case one can lift the $\SO(n)$ bundle to a $\Spin(n)$ bundle and, in particular, different lifts differ by an element in $H^1(\cM;\bZ_2)$.\footnote{Elements of the cohomology $H^1(\cM;\bZ_2)$ act freely and transitively on the spin structures of $\cM$ and hence it classify them.} Intuitively, this means that the number of spin structures equals the number of non-trivial 1-cycles in $\cM$. For example, a genus $g$ Riemann surface with $n$ punctures has $2^{2g+n}$ distinct spin structures. Given a spin structure on a manifold $\cM$, one can ask if there are invariant Killing spinors. Such Killing spinors are solutions to the Killing equation,
\begin{equation}\label{Killing}
	\nabla_X \psi = -\alpha X\cdot \psi\,,
\end{equation}
where $X$ is a tangent vector and $\cdot$ represents Clifford multiplication. The constant $\alpha$ is determined in terms of the Ricci scalar of the manifold. If it is Ricci flat, $\alpha=0$ and the Killing spinor is often called a parallel spinor. More generally, for $\alpha \neq 0$ such spinors are called conformal Killing spinors \cite{Bar:1993gpi}. 

In this work we study a class of $\AdS_4\times \cM_7$ backgrounds of eleven-dimensional supergravity. In particular, we are interested in manifolds $\cM_7$ which have multiple spin structures preserving some number of invariant Killing spinors. The importance of specifying the spin structure as defining data for the supergravity background was highlighted in \cite{Figueroa-OFarrill:2005vxy}. We review the parts of their discussion that we find relevant in the context of the holographic applications of interest here. Since AdS$_4$ is maximally symmetric and admits invariant spinors, the supersymmetry preserved by the supergravity background --- or, equivalently, by the dual field theory --- is determined by the number of Killing spinors on the transverse space, $S^7/\bZ_k$.  Our focus here is on orbifolds where the action by $\bZ_k$ is isometric and free so that the quotient is smooth and locally isometric to $S^7$; in other words, the 7d manifold $\cM_7$ is a lens space. 

The spin structures on $S^7/\bZ_k$ are given by the possible lifts of the action $\bZ_k \subset \SO(8)$ to the spin bundle. Such lifts exist if and only if there is a subgroup $\hat{\Gamma}\subset \Spin(8)$ which is mapped isometrically to $\bZ_k$ under the covering map. Having specified a specific lift of $\bZ_k$ we can investigate whether $S^7/\bZ_k$ admits Killing spinors. By B\"ar's cone construction \cite{Bar:1993gpi}, Killing spinors on $S^7/\bZ_k$ are in one-to-one correspondence with parallel spinors on $\bC^4/\bZ_k$. In other words, we can identify $S^7/\bZ_k$ with the unit sphere in $\bC^4/\bZ_k$ with the orbifold action induced from a linear representation of $\bZ_k$ on $\bC^4$. To be concrete let us denote the generator of $\bZ_k$ by $a$, such that $a^k=1$ and its action on $\bC^4$, parameterized by the complex variables $z_{1,2,3,4}$, is given by\footnote{Here and in the following we slightly abuse notation by identifying the group element $a$ with its action as a linear representation on $\bC^4$.}
\begin{equation}\label{orbifoldaction}
	a\,:\,  \bC^4\rightarrow \bC^4\,:\, \left(z_1 , z_2 , z_3 , z_4\right)\mapsto \left( \e^{2\pi i/k}z_1,\e^{2\pi i n_1 /k}z_2,\e^{2\pi i n_2 /k}z_3,\e^{2\pi i n_3 /k} z_4\right)\,,
\end{equation}
where $n_i \in \bN$ and, in order to ensure the smoothness of the quotient we impose that all $n_i$ are coprime to $k$, i.e.
\begin{equation}
	\gcd(n_i,k)=1\,.
\end{equation}
Without loss of generality we choose $1\leq n_1 \leq n_2 \leq n_3 < k$. One can now show that the most general lift of $a$ to $\Spin(8)\in \Cliff(8)$ is given by
\begin{equation}\label{spinsign}
	\hat{a} = \epsilon \exp \left[\f{\pi}{k}\left(\gamma_{12}+ n_1 \gamma_{34}+ n_2 \gamma_{56}+ n_3 \gamma_{78}\right)\right]\,,
\end{equation}
where in order to obey $\hat{a}^k = \epsilon^k (-1)^{1+n_1+n_2+n_3} \mathbf{1}$ we have to impose $\epsilon=\pm 1$. 

When $k$ is even, $n_1$, $n_2$ and $n_3$ are odd, and hence $1+n_1+n_2+n_3$ is even. In this case we have that $\hat{a}^k=1$ for both $\epsilon=\pm1$ and therefore there are two inequivalent spin structures. On the other hand, for odd $k$ we have to choose $\epsilon=(-1)^{1+n_1+n_2+n_3}$ in order to satisfy $\hat{a}^k=1$ and thus there is a unique spin structure for the quotient manifold. Indeed, this observation is in line with the fact that for odd $k$ we have $H^1(S^7/\bZ_k,\bZ_2)=0$, while for even $k$ $H^1(S^7/\bZ_k,\bZ_2)=\bZ_2$.\footnote{These cohomology groups can be easily computed using the universal coefficient theorem which states that $H^1(\cM,\bZ_2)=\Hom(\pi_1(\cM),\bZ_2)$ which in our case simply implies $H^1(S^7/G,\bZ_2)=\Hom(\pi_1(S^7/G),\bZ_2)=\Hom(G,\bZ_2)$.}

Once we have chosen a particular spin structure, we need to ask how many Killing spinors are preserved on $S^7/\bZ_k$. The round seven-sphere has a single spin structure and has the maximal number of 8 Killing spinors. To determine the number of Killing spinors on the lens spaces of interest we have to select the spinors invariant under the $\bZ_k$ action. Since $\gamma_i^2=-1$, its eigenvalues are $\pm i$ and therefore the eigenvalues of $\hat{a}$ are given by 
\begin{equation}
	\epsilon \exp\left[\f{i\pi}{k}\left(\sigma_0+n_1 \sigma_1+ n_2 \sigma_2+ n_3 \sigma_3  \right)\right]\,,
\end{equation}
with signs $\sigma_i = \pm 1$. Depending on the value of $\epsilon$ this is equal to unity if and only if
\begin{align}
	\sigma_0+n_1\sigma_1+n_2\sigma_2+n_3\sigma_3&=0\,,\pm 2k\,, & \text{for}&\quad \epsilon=1\,, \label{poseigenvalues}\\
	\sigma_0+n_1\sigma_1+n_2\sigma_2+n_3\sigma_3&=\pm k\,, & \text{for}&\quad \epsilon=-1\,.\label{negeigenvalues}
\end{align}

The main example we study below is given by the orbifold with the action determined by the integers $n_1=n_2=n_3=1$. The resulting supergravity background is very well known. Upon choosing the periodic spin structure, i.e. $\epsilon=1$, the dual field theory is given by the $\UU(N)_{k} \times \UU(N)_{-k}$ $\cN=6$ ABJM SCFT \cite{Aharony:1999ti}. This is indeed in agreement with the analysis above, as in this case \eqref{poseigenvalues} always has at least six solutions by choosing two of the $\sigma_i$ positive and two negative. When $k=2$ there are two additional solutions given by all $\sigma$'s either positive or negative. For $k=1$ we have simply the round $S^7$ which has 8 Killing spinors. Again, this matches perfectly with the expectation from the ABJM theory which has $\mathcal{N}=6$ supersymmetry for $k>2$, which is enhanced to $\mathcal{N}=8$ supersymmetry for $k=1,2$ due to the presence of low-dimension monopole operators \cite{Bashkirov:2010kz}. 

For the antiperiodic spin structure, i.e. $\epsilon=-1$, we find that the condition on the eigenvalues \eqref{negeigenvalues} is much more restrictive. In fact for generic $k$ there are no solutions to \eqref{negeigenvalues} and thus no invariant Killing spinors. Two special cases are given by $k=2$ and $k=4$. For $k=2$, we can find again 8 solutions corresponding to one $\sigma$ positive/negative and the three others having the opposite sign. In this case the resulting geometry is given by $S^7/\bZ_2 = \bR\bP^7$ which indeed is the unique lens space with the maximal number of invariant Killing spinors for both spin structures \cite{FRANC1987277}. In this case both spin structures should give rise to the same holographically dual SCFT, i.e. the ABJM theory at level $k=2$. The case $k=4$ is more interesting since for the antiperiodic spin structure there are only 2 solutions to \eqref{negeigenvalues} corresponding to all $\sigma$'s either positive or negative. We thus find that $S^7/\bZ_4$ preserves a total of 8 invariant Killing spinors, 6 for the periodic spin structure and 2 for the antiperiodic one.  This result suggests that for $k=4$ and $n_{1,2,3}=1$ the antiperiodic spin structure leads to an AdS$_4\times S^7/\bZ_4$ solution of 11d supergravity which is holographically dual to a new 3d $\cN=2$ SCFT, distinct from the $k=4$ ABJM theory with $\mathcal{N}=6$ supersymmetry. In Section~\ref{sec:KK} we support this claim by studying the KK spectrum of 11d supergravity on $S^7/\bZ_4$.

The existence of two different supersymmetry preserving spin structures can also be deduced from the geometry of the AdS$_4\times S^7/\bZ_4$ solution of 11d supergravity. Consider a Freund-Rubin solution in 11d of the form
\begin{equation}\label{eq:FRsoln}
\begin{split}
	\ds_{11}^2 &= R^2 \left( \f14 \ds_{\AdS_4}^2 +\ds^2_{\rm KE}+ (\dd\xi + A_{\rm KE})^2  \right)\,,\\
	G_4 &= \f{3R^3}{8}\dvol_{\AdS_4}\,.
\end{split}
\end{equation}
where $\ds^2_{\rm KE}$ is a smooth K\"ahler-Einstein manifold of real dimension 6 and $\dd A_{\rm KE}$ is its K\"ahler form. The 7d internal space is then a regular Sasaki-Einstein manifold provided that the coordinate $\xi$ has period $\pi/2$.\footnote{We use the normalization $4\dd A_{\rm KE}=\mathcal{R}_{\rm KE}$ where $\mathcal{R}_{\rm KE}$ is the Ricci two-form for the K\"ahler-Einstein metric $\ds^2_{T}$.} Sasaki-Einstein manifolds admit two Killing spinors that have equal and opposite charges under the Reeb vector $\partial_{\xi}$, see \cite{Sparks:2010sn} for a review. If one takes the period of $\xi$ to be smaller than $\pi/2$ then the Killing spinors are not globally well-defined and supersymmetry is broken. Therefore, any $\bZ_q$ orbifold (for $q>1$) of a general Sasaki-Einstein manifold along the Reeb vector is not supersymmetric. In some cases however $\xi$ can have a period larger than $\pi/2$. This is precisely what happens when the K\"ahler-Einstein space is $\mathbf{CP}^3$. Then if the period $\xi$ is $2\pi$ one finds the metric on the round unit volume $S^7$ written as a Hopf fibration. In this case there are of course 8 Killing spinors on the internal manifold and the background in \eqref{eq:FRsoln} is maximally supersymmetric. The $\bZ_k$ orbifold of $S^7$ in \eqref{orbifoldaction} with $n_{1,2,3}=1$ acts precisely on the coordinate $\xi$. We therefore conclude that for general $k>1$ only 6 of the 8 Killing spinors on $S^7$ are preserved. The cases $k=2$ and $k=4$ are however special. For $k=2$ we have $S^7/\bZ_2=\mathbf{RP}^7$ which admits 8 Killing spinors. For $k=4$, on the other hand, $\xi$ has period $2\pi/4=\pi/2$. This is precisely the correct period for the 7d metric in \eqref{eq:FRsoln} to have the form of a regular Sasaki-Einstein metric with 2 Killing spinors. We therefore conclude that the AdS$_4\times S^7/\bZ_4$ 11d supergravity solution has the 6 ``standard'' Killing spinors that exists for all $\bZ_k$ orbifolds of $S^7$ along $\xi$, together with 2 ``extra'' Killing spinors. Importantly however, as discussed above, these two sets of spinors are not compatible with the same spin structure, i.e. they belong to different spin bundles. A more detailed discussion on this point of view on the two spin structures on $S^7/\bZ_4$ can be found in \cite{Martelli:2012sz}.

\subsection{Generalizations}
\label{subsec:generalization1}

In addition to the main example discussed in detail above one can also consider other smooth orbifold actions of the type defined in \eqref{orbifoldaction}. As emphasized in \cite{Figueroa-OFarrill:2005vxy}, for any $k=4p$ with $p \in \mathbf{N}$ there are two choices of spin structure preserving some amount of invariant Killing spinors for particular choices of $n_{1,2,3}$. Indeed, choosing the orbifold action defined by any of the following parameters
\begin{align}
	(n_1,n_2,n_3) &= (1,2p-1,2p-1)\,,\label{first}\\
	(n_1,n_2,n_3) &= (1,2p+1,2p+1)\,,\label{second}\\
	(n_1,n_2,n_3) &= (1,2p-1,2p+1)\,,\label{third}\\
	(n_1,n_2,n_3) &= (2p-1,2p+1,4p-1)\,,\label{fourth}
\end{align}
one can show that the resulting lens spaces have two different spin structures with invariant Killing spinors. For the periodic spin structure in \eqref{poseigenvalues} one finds 4 invariant Killing spinors while for the antiperiodic spin structure \eqref{poseigenvalues} there are 2 invariant spinors. The holographic interpretation of this is that the corresponding AdS$_4 \times S^7/\bZ_k$ supergravity solution is dual to an $\mathcal{N}=4$ SCFT for the periodic spin structure and an $\mathcal{N}=2$ SCFT for the antiperiodic one.\footnote{The $\mathcal{N}=4$ SCFTs for some of these orbifold action were studied in \cite{Terashima:2008ba}.} We provide more evidence for this expectation by studying the KK spectrum of 11d supergravity on such internal manifolds for both choices of spin structure in Section~\ref{subsec:generalization2}. 

\section{KK spectrum of 11d supergravity on AdS$\mathbf{_4\times S^7/\bZ_k}$}
\label{sec:KK}

The different choices of spin structures on $S^7/\bZ_k$ suggest the existence of different 3d SCFTs with varying amount of supersymmetry. To better understand their properties we study the KK spectrum of 11d supergravity on AdS$_4\times S^7/\bZ_k$ by paying particular attention to the effects of the different spin structures. Since the dual QFT is conformal and supersymmetric the KK modes should be dual to operators that organize into superconformal multiplets. We show below that depending on the value of $k$ and the choice of spin structure the spectrum of operators is different and organizes into $\mathcal{N}=8$, $\cN=6$, $\cN=4$, or $\cN=2$ superconformal multiplets. To illustrate our results in more detail we focus on $S^7/\bZ_4$ and show that the KK spectrum for the two spin structures organizes into either $\mathcal{N}=6$ or $\mathcal{N}=2$ superconformal multiplets.

To construct the KK spectrum on the lens space $S^7/\bZ_4$ we start from the KK spectrum on the round seven sphere and carefully determine which modes are invariant under the orbifold action. The field theory dual to $\AdS_4\times S^7$ is the $k=1$ $\UU(N)\times \UU(N)$ ABJM theory which preserves $\cN=8$ supersymmetry. The spectrum of $\cN=8$ multiplets obtained after the KK reduction on the round $S^7$ was worked out in \cite{Duff:1986hr,Biran:1983iy}, see also \cite{DHoker:2000pvz} for a review in the holographic context, and this is the starting point of our analysis.

The 3d $\cN=8$ superconformal algebra is $\osp(8|4)$ and contains an $\so(8)$ $R$-symmetry. We label $R$-symmetry representations by their Dynkin labels $(\alpha_1,\alpha_2,\alpha_3,\alpha_4)$. Our conventions for the Dynkin labels and $\SO(8)$ triality frame are presented in Appendix~\ref{app:group}. Here we simply note that the 16 Poincar\'e supercharges transform in the $\SO(8)$ representation\footnote{We use the notation introduced in \cite{Cordova:2016emh} to label superconformal multiplets. $[j]^{(r)}_{\Delta}$ labels the representation with Lorentz $\su(2)$ Dynkin label $j$, $R$-symmetry representation $(r)$ and conformal dimension $\Delta$. More details on our notation and conventions are given in Appendix \ref{app:group}.}
\begin{equation}
	Q_\alpha \in [1]^{(0,0,0,1)}_{\f12}\,,
\end{equation}
which for our choice of triality frame corresponds to the $\mathbf{8}_s$ representation. 

The $\cN=8$ superconformal multiplets arising from the KK modes on the round sphere have maximal spin $2$, or equivalently $j=4$, and the KK levels are labeled by a single integer $n$. At each level, the fluctuations fall into representations of the $\cN=8$ superconformal group and moreover all superconformal multiplets are short. Level $n=0$ corresponds to the massless $\cN=8$ supergravity multiplet and forms the $B_1[0]_1^{(2,0,0,0)}$ $\cN=8$ superconformal multiplet which contains the energy momentum tensor and the $\SO(8)$ R-symmetry current. The higher levels $n>0$ correspond to massive multiplets and level by level organize into the $B_1[0]_{\f{n}{2}+1}^{(n+2,0,0,0)}$ superconformal multiplet. The supergravity modes comprising the supermultiplets from level $n=0$ and $n\geq 1$ are presented in Table~\ref{tab:multipletn0} and Table~\ref{tab:multipletn}, respectively. 
\begin{table}[!htb]
	\centering
	\begin{tabular}{c|c|c|c|c}
		$j$ & Field & $\SO(8)$ irrep & $\SO(8)$ Dynkin labels & $\Delta$\\
		\hline
		$4$ & $e^{(0)}_\mu{}^a$ & $\mathbf{1}$ & $(0,0,0,0)$ & $3$\\
		$3$ & $\psi^{(0)}_\mu{}^I$ & $\mathbf{8}_s$ & $(0,0,0,1)$ & $\f52$\\
		$2$ & $A^{(0)}_\mu{}^{IJ}$ & $\mathbf{28}$ & $(0,1,0,0)$ & $2$\\
		$1$ & $\chi^{(0)IJK}$ & $\mathbf{56}_s$ & $(1,0,1,0)$ & $\f32$\\
		$0_+$ & $S^{(0)[IJKL]_+}$ & $\mathbf{35}_v$ & $(2,0,0,0)$ & $1$\\
		$0_-$ &$P^{(0)[IJKL]_-}$& $\mathbf{35}_c$ & $(0,0,2,0)$ & $2$\\
	\end{tabular}
\caption{The massless $\cN=8$ supermultiplet $B_1[0]_1^{(2,0,0,0)}$. We have indicated the supergravity modes that comprise the multiplet, their $\SO(8)$ representation, as well as the conformal dimension $\Delta$ of the operator in the dual SCFT.}
\label{tab:multipletn0}
\end{table}
\begin{table}[!htb]
	\centering
	\begin{tabular}{c|c|c|c}
		$j$ & Field & $\SO(8)$ Dynkin labels& $\Delta$\\
		\hline
		$4$ & $e^{(n)}_\mu{}^a$ & $(n,0,0,0)_{n\geq 0}$ & $\f{n}{2}+3$\\
		\hline
		\multirow{2}{*}{$3$} & \multirow{2}{*}{$\psi^{(n)}_\mu{}^I$} & $(n,0,0,1)_{n\geq 0}$ & $\f{n}{2}+\f52$\Tstrut\\
		&& $(n-1,0,1,0)_{n\geq 1}$ & $\f{n}{2}+\f72$\Tstrut\\
		\hline
		\multirow{3}{*}{$2$} & \multirow{3}{*}{$A^{(n)}_\mu{}^{IJ}$} & $(n,1,0,0)_{n\geq 0}$ & $\f{n}{2}+2$\Tstrut \\
		&& $(n-1,0,1,1)_{n\geq 1}$ &  $\f{n}{2}+3$\Tstrut\\
		&& $(n-2,1,0,0)_{n\geq 2}$ &  $\f{n}{2}+4$\Tstrut\\
		\hline
		\multirow{4}{*}{$1$} & \multirow{4}{*}{$\chi^{(n)IJK}$} & $(n+1,0,1,0)_{n\geq 0}$ &  $\f{n}{2}+\f32$\Tstrut \\
		&& $(n-1,1,1,0)_{n\geq 1}$ & $\f{n}{2}+\f52$\Tstrut\\
		&& $(n-2,1,0,1)_{n\geq 2}$ & $\f{n}{2}+\f72$\Tstrut \\
		&& $(n-2,0,0,1)_{n\geq 2}$ & $\f{n}{2}+\f92$\Tstrut \\
		\hline
		\multirow{3}{*}{$0$} & \multirow{3}{*}{$S^{(n)[IJKL]_+}$} & $(n+2,0,0,0)_{n\geq 0}$ & $\f{n}{2}+1$\Tstrut\\
		&& $(n-2,2,0,0)_{n\geq 2}$ & $\f{n}{2}+3$\Tstrut \\
		&& $(n-2,0,0,0)_{n\geq 2}$ & $\f{n}{2}+5$ \Tstrut\\
		\hline
		\multirow{2}{*}{$0$} & \multirow{2}{*}{$P^{(n)[IJKL]_-}$} & $(n,0,2,0)_{n\geq 0}$ & $\f{n}{2}+2$\Tstrut \\
		&& $(n-2,0,0,2)_{n\geq 2}$ &$\f{n}{2}+4\Tstrut$
	\end{tabular}
	\caption{The massive $\cN=8$ supermultiplets $B_1[0]_{\f{n}{2}+1}^{(n+2,0,0,0)}$ at KK level $n$. The notation is the same as in Table~\ref{tab:multipletn0}.}
	\label{tab:multipletn}
\end{table}

Now that we have described the full KK spectrum of the round seven-sphere, we have to determine which part of the spectrum remains invariant under the orbifold action \eqref{orbifoldaction} with a given spin structure. We again resort to our main example of $S^7/\bZ_4$ to illustrate this but will comment on its various generalizations at the end of this section. For $S^7/\bZ_4$ the orbifold action is defined by the three integers $n_{1,2,3}=1$ in \eqref{orbifoldaction} and manifestly preserves an $\SU(4)\times \UU(1)\subset \SO(8)$ isometry.

As discussed above, upon choosing the periodic spin structure this orbifold corresponds to the $\cN=6$ ABJM theory at level $k=4$. Choosing the antiperiodic spin structure on the other hand should result in a different $\cN=2$ SCFT. In the former case the R-symmetry is given by $\SU(4)$ while the $\UU(1)$ is a flavor symmetry. In the latter case on the other hand the $\UU(1)$ is the superconformal R-symmetry of the $\cN=2$ theory while the $\SU(4)$ symmetry is a flavor symmetry. In order for these two possibilities to coexist, the KK spectrum of $S^7/\bZ_4$, with the above orbifold action, should organize in multiplets of both the $\cN=6$ superalgebra $\osp(6|4)$ and the $\cN=2$ superalgebra $\osp(2|4)$. This is non-trivial and we show how it is realized in detail below.

To find the KK spectrum on $S^7/\bZ_k$ we proceed in two steps. First, we decompose the spectrum under the branching 
\begin{equation}\label{eq:SO8branch}
\SO(8)\rightarrow \SU(4)\times \UU(1)\,.
\end{equation}
In the resulting expressions we normalize the $\UU(1)$ charges such that
\begin{align}
	\mathbf{8}_v &\rightarrow \mathbf{4}_{\f14} \oplus \mathbf{\bar{4}}_{-\f14}\,,\\
	\mathbf{8}_s &\rightarrow \mathbf{6}_0 \oplus \mathbf{1}_{\f12} \oplus \mathbf{1}_{-\f12}\label{Qbranch}\,.
\end{align}
This choice is such that the $\SU(4)$ singlets in \eqref{Qbranch} have the canonical $\UU(1)$ charges to become a $\UU(1)$ R-symmetry. Next, we have to select those states that are invariant under the $\bZ_k$ orbifold action \eqref{orbifoldaction}. For the bosonic sector the answer is straightforward as we can simply select the KK modes that are stable under the orbifold action. In our normalization this corresponds to the ones with $\UU(1)$ charge divisible by $k/4$.

It is somewhat more subtle to select the correct fermionic modes invariant under the orbifold action. The choice of spin structure determines the periodicity of spinors around non-trivial 1-cycles in the geometry. Indeed, unlike $S^7$, for even $k$, $S^7/\bZ_k$ is no longer simply connected and one has to choose the periodicity of the fermions around the non-contractible cycle. Writing the sphere as a Hopf fibration $S^1 \hookrightarrow S^7\rightarrow \bC\bP^3$ the orbifold acts on the Hopf fiber and the choice of spin structure determines the periodicity of spinors along this fiber. The spin structure with positive sign $\epsilon=1$ in \eqref{spinsign} corresponds to periodic fermions while the sign for $\epsilon=-1$ corresponds to antiperiodic spinors. For this reason we find that for the periodic spin structure we can follow the same rule as for the bosonic spectrum and select fermionic modes with $\UU(1)$ charge divisible by $k/4$. For the antiperiodic spin structure on the other hand the invariant fermionic spectrum is obtained by a different rule and we have to retain the fermionic modes with $\UU(1)$ charges equal to $q= \f{k}{8}+\f{mk}{4}$, $m\in \bZ$. 

Inspecting the branching rule \eqref{Qbranch}, we see that for the periodic spin structure we always preserve the supercharges transforming in the $\mathbf{6}_0$ while for the antiperiodic spin structure for generic $k$ we do not preserve any of the supercharges. Only for $k=4$ the additional two supercharges have the compatible $\UU(1)$ charges and hence in this case we preserve the supercharges transforming in the representations $\mathbf{1}_{\f12}\oplus\mathbf{1}_{-\f12}$. The discussion above leads to the following simple rule to find the spectrum on $S^7/\bZ_4$ 

\noindent \textbf{The $\bZ_4$ rule:} \textit{If we choose the periodic spin structure the $\bZ_4$ invariant KK modes, both bosons and fermions, are those with integer $\UU(1)$ charge. For the antiperiodic spin structure, the invariant KK modes are the ones with integer $\UU(1)$ charge for the bosons and half-integer charge for the fermions.}

With this rule at hand, the task of finding the KK spectrum on $S^7/\bZ_4$ for both choices of spin structure becomes a straightforward group theory exercise.

\subsection{Periodic spin structure}
\label{subsec:periodZ4}

When we choose the periodic spin structure, the corresponding 3d SCFT preserves $\cN=6$ superconformal symmetry. The superconformal algebra is given by $\osp(6|4)$, hence the R-symmetry is $\SO(6)_R\simeq \SU(4)_R$ and the $12$ Poincar\'e supercharges transform in the $\so(6)$ vector representation of the $R$-symmetry. In order for the KK spectrum to correspond to this $\cN=6$ theory it should organize into $\cN=6$ supermultiplets. Such multiplets can be either long or short, depending on whether there are any null states in the multiplet. As in \cite{Cordova:2016emh} we indicate this information by using $L$ for the long multiplets and $A_{1,2}$ or $B_{1,2}$ for the various shortening conditions. In addition, all the fields in a supermultiplet should transform identically under the $\UU(1)$ flavor symmetry in \eqref{eq:SO8branch} and thus should have the same $\UU(1)$ charge.

To illustrate this let us first consider the $\bZ_4$ invariant part of the KK spectrum of $S^7$. At each KK level the spectrum organizes into $\cN=6$ multiplets so we can consider them level by level. The invariant modes at the lowest level $n=0$ are given in Table~\ref{tab:KKZ4n0}. Note that all the modes are neutral under the $\UU(1)$ flavor symmetry. These modes are precisely the ones that are dual to the operators in the $\cN=6$ energy momentum tensor multiplet $B_1[0]_1^{(1,0,1)}$. We can proceed similarly for the higher KK levels. In particular, at level $n=1$ we find that there are no $\bZ_4$ invariant KK modes and as shown in Appendix~\ref{app:branching}, this continues to hold for all odd KK levels. The next interesting level is therefore $n=2$ whose invariant spectrum is given in Table~\ref{tab:KKZ4n2} in Appendix \ref{app:moretables}. In this case the we find representations of $\UU(1)$ flavor charge $0$ and $\pm 1$. The charge $\pm 1$ representations organize into a pair of $B_1[0]_2^{(4,0,0)}$ and $B_1[0]_2^{(0,0,4)}$ multiplets while the $\UU(1)$ neutral part forms a $B_1[0]_2^{(2,0,2)}$ multiplet. The higher even levels, $n>2$, organize in a similar way into $\cN=6$ multiplets, see \cite{Liu:2016dau} for a recent discussion. In particular, each $\cN=8$ supermultiplet can be decomposed into a set of $\cN=6$ multiplets as follows,
\begin{equation}\label{N6multiplets}
	\begin{aligned}
		B_1[0]_{\f{n}{2}+1}^{(n+2,0,0,0)}\rightarrow& \left(-\tfrac{n+2}{4} \right) B_1[0]^{(0,0,n+2)}_{\f{n}{2}+1} + \left(\tfrac{n+2}{4} \right)B_1[0]^{(n+2,0,0)}_{\f{n}{2}+1}\\
		&+\sum_{r=0}^n \left(\tfrac{2r-n}{4} \right)B_1[0]_{\f{n}{2}+1}^{(1+r,0,n+1-r)}\,.
	\end{aligned}
\end{equation}
In this equation, the coefficient in front of each supermultiplet indicates the $\UU(1)$ flavor charge and to obtain the correct $\bZ_4$ invariant spectrum we should only select terms on the right hand side that have integer $\UU(1)$ flavor charge.

\begin{table}[!htb]
	\centering
	\begin{tabular}{c|c|c|c}
		Field & $\SU(4)\times \UU(1)$ irrep & Lorentz $\su(2)$ Dynkin label $j$ &$\Delta$ \\
		\hline
		$e^{(0)}_\mu{}^a$ & $\mathbf{1}_0$ & 4 & $3$\\
		$\psi^{(0)}_\mu$ & $\mathbf{6}_0$& 3 & $\f52$\\
		$A^{(0)}_\mu$ & $\mathbf{15}_0 + \mathbf{1}_0$& 2 & $2$ \\
		$\chi^{(0)}$ & $\mathbf{10}_0+\overline{\mathbf{10}}_0 + \mathbf{6}_0$& 1 & $\f32$\\
		$S^{(0)}$ & $\mathbf{15}_0$& 0 & $1$\\
		$P^{(0)}$ & $\mathbf{15}_0$& 0 & $2$\\
	\end{tabular}
	\caption{The KK spectrum for the level $n=0$ of 11d supergravity on AdS$_4\times S^7/\bZ_4$ with the periodic spin structure. The modes organize into the $\mathcal{N}=6$ energy momentum tensor multiplet of the dual SCFT.}
	\label{tab:KKZ4n0}
\end{table}

It is also instructive to rewrite the $\cN=6$ multiplets described above in $\cN=2$ language. This can be done by further breaking the $R$-symmetry as $\SU(4) \rightarrow \SU(2)\times \SU(2)\times \UU(1)_{\cN=6}$. The $\UU(1)_{\cN=6}$ factor plays the role of the $\cN=2$ $R$-symmetry while the $\SU(2)\times \SU(2)$ symmetry is an additional flavor symmetry from the $\cN=2$ point of view. To illustrate the decomposition into $\cN=2$ multiplets consider the level $n=0$ KK modes. After breaking the symmetries we can reorganize them into the following $\cN=2$ multiplets:
\begin{equation}
	\begin{aligned}
		B_1[0]_{1}^{(2,0,0,0)}\rightarrow & (\mathbf{1},\mathbf{1})_0 \left(A_2\overline{A}_2[0]_1^{(0)}+A_1\overline{A}_1[2]_2^{(0)}\right) + \left((\mathbf{3},\mathbf{1})_0+(\mathbf{1},\mathbf{3})_0\right) A_2\overline{A}_2[0]_1^{(0)} \\
		&+ (\mathbf{2},\mathbf{2})_0 \left(L\overline{B}_1[0]_1^{(1)}+B_1\overline{L}[0]_1^{(-1)}+A_1\overline{A}_1[1]_{\f32}^{(0)}\right)\,. 
	\end{aligned}
\end{equation}
The representations in front of each multiplet indicate the transformation properties under the $\SU(2)\times \SU(2)$ flavor symmetry and the charge under the $\UU(1)$ in \eqref{eq:SO8branch}. We recognize the appearance of the $\cN=2$ energy momentum tensor multiplet $A_1\overline{A}_1[2]_2^{(0)}$, a flavor current multiplet $A_2\overline{A}_2[0]_1^{(0)}$ corresponding to the $\UU(1)$ in \eqref{eq:SO8branch}, as well as flavor current multiplets for the $\SU(2)\times \SU(2)$ symmetry. Importantly, we also have the supersymmetry current multiplet $A_1\overline{A}_1[1]_{\f32}^{(0)}$ indicating that this theory contains a larger supersymmetry algebra.

\subsection{Anti-periodic spin structure}
\label{subsec:antiperZ4}

After discussing the periodic spin structure we now turn our attention to the more intriguing case of the antiperiodic spin structure. For this choice of spin structure on $S^7/\bZ_4$ the holographic dual 3d SCFT preserves the $\cN=2$ superconformal algebra $\osp(2|4)$, which has R-symmetry $\UU(1)_R$. In our conventions the two supercharges have $R$-charge $\pm \f12$ and can be identified with the two $\SU(4)$ singlets in \eqref{Qbranch}. Choosing the antiperiodic spin structure does not break any of the continuous global symmetries in the problem and in particular the $\SU(4)$ symmetry in this case becomes a flavor symmetry. The KK spectrum on AdS$\times S^7/\bZ_4$ organizes into supermultiplets but this time with respect to the $\cN=2$ algebra. All operators in a given superconformal multiplet transform identically under the $\SU(4)$ flavor symmetry.

The spectrum of bosonic operators is identical to the one for the periodic spin structure, but the fermionic spectrum differs and includes only half integer $\UU(1)$ charges. Indeed, this is crucial in order to be able to arrange the KK spectrum into $\cN=2$ representations. Again, before discussing the general case, we illustrate how this works for the low-lying KK levels $n=0,2$. Following the $\bZ_4$ rule, we find the $\bZ_4$ invariant spectrum at level $n=0$ as tabulated in Table~\ref{tab:KKZ4altn0}. Similarly, the spectrum at level $n=2$ is given in Table~\ref{tab:KKZ4altn2} in Appendix~\ref{app:moretables}.

\begin{table}[!htb]
	\centering
	\begin{tabular}{c|c|c|c}
		Field & $\SU(4)\times \UU(1)$ irrep & Lorentz $\su(2)$ Dynkin label $j$ &$\Delta$ \\
		\hline
		$e^{(0)}_\mu{}^a$ & $\mathbf{1}_0$ & 4 & $3$\\
		$\psi^{(0)}_\mu$ & $\mathbf{1}_{\f12}+\mathbf{1}_{-\f12}$& 3 & $\f52$\\
		$A^{(0)}_\mu$ & $\mathbf{15}_0 + \mathbf{1}_0$& 2 & $2$ \\
		$\chi^{(0)}$ & $\mathbf{15}_{\f12}+\mathbf{15}_{-\f12}$& 1 & $\f32$\\
		$S^{(0)}$ & $\mathbf{15}_0$& 0 & $1$\\
		$P^{(0)}$ & $\mathbf{15}_0$& 0 & $2$\\
	\end{tabular}
	\caption{The KK spectrum for the level $n=0$ of 11d supergravity on AdS$_4\times S^7/\bZ_4$ with the antiperiodic spin structure.}
	\label{tab:KKZ4altn0}
\end{table}

As expected, the spectrum nicely organizes into $\cN=2$ multiplets and it does so level by level. At level $n=0$ the fields in the $\SU(4)$ singlet representation, $\mathbf{1}$, form the $\cN=2$ stress tensor multiplet $A_1\overline{A}_1[2]_2^{(0)}$ which also contains the $\UU(1)$ R-symmetry current. The modes transforming in the $\mathbf{15}$ representation form an $\SU(4)$ flavor current multiplet $A_2\overline{A}_2[0]_1^{(0)}$. Note that in this case there is no additional SUSY current multiplet since we have an honest $\cN=2$ SCFT, distinct from the usual $\mathcal{N}=6$ ABJM theory. At level $n=2$ the resulting supermultiplets are given in Table~\ref{tab:n2altmultiplet} where notably in addition to short multiplets we also find a number of long multiplets.

\begin{table}[!htb]
	\centering
	\begin{tabular}{c|c|c}
		$\SU(4)$ Dynkin label & $\SU(4)$ irrep & $\cN=2$ multiplet\\
		\hline
		$(0,0,0)$ & $\mathbf{1}$ &  $L\overline{L}[0]_4^{(0)}$ \Tstrut\\
		\hline
		$(1,0,1)$ & $\mathbf{15}$ & $L\overline{L}[2]_3^{(0)}$ \Tstrut\\
		\hline
		$(0,2,0)$ & $\mathbf{20}^\prime$ & $L\overline{L}[0]_3^{(0)}$\Tstrut\\
		\hline
		$(4,0,0)$ & $\mathbf{35}$  & $L\overline{B}_1 [0]_2^{(1)}$ \Tstrut\\
		\hline
		$(0,0,4)$ & $\overline{\mathbf{35}}$ &  $B_1\overline{L} [0]_2^{(-1)}$ \Tstrut\\
		\hline
		$(2,1,0)$ & $\mathbf{45}$ &  $L\overline{A}_1[1]_{\f52}^{(\f12)}$\Tstrut\\
		\hline
		$(0,1,2)$ & $\overline{\mathbf{45}}$ &  $A_1\overline{L}[1]_{\f52}^{(-\f12)}$ \Tstrut\\
		\hline
		$(2,0,2)$ & $\mathbf{84}$ & $L\overline{L}[0]_2^{(0)}$ \Tstrut
	\end{tabular}
	\caption{The $\cN=2$ multiplets from KK level $n=2$ for the antiperiodic spin structure.}
	\label{tab:n2altmultiplet}
\end{table}

We can repeat the same exercise for higher KK levels as is illustrated in Appendix~\ref{app:moretables} where we explicitly organize the modes in the first four KK levels into $\mathcal{N}=2$ multiplets. One can furthermore show in general that, level by level, the KK modes organize into $\cN=2$ superconformal multiplets where for level $n\geq2$ the result is given by
\begin{align}\label{N2multiplets}
		B_1[0]_{\f{n}{2}+1}^{(n+2,0,0,0)} \rightarrow & \sum_{l=-\lfloor\f{n+2}{4}\rfloor}^{\lfloor\f{n+2}{4}\rfloor}\sum_{i=0}^2 \left(\f{n}{2}-1+i+2l,0,\f{n}{2}-1+i-2l\right)X\overline{Y}[(2i)\hspace{-8pt}\mod 4]_{\f{n}{2}-i+3}^{(l)}\nn\\
		&+ \sum_{l=-\lfloor\f{n+2}{4}\rfloor}^{\lfloor\f{n-2}{4}\rfloor}\sum_{i=0}^1 \left(\f{n}{2}+i+2l,1,\f{n}{2}-2+i-2l\right)X\overline{Y}[1]_{\f{n}{2}-i+\f52}^{\left(l+\f12\right)} \\ 
		&+ \sum_{l=-\lfloor\f{n-2}{4}\rfloor}^{\lfloor\f{n-2}{4}\rfloor} \left(\f{n}{2}-1+2l,2,\f{n}{2}-1-2l\right)L\overline{L}[0]_{\f{n}{2}+2}^{(l)} \,. \nn
\end{align}
A few comments about this formula are in order. The three numbers in the parenthesis $(\boldsymbol{\cdot},\boldsymbol{\cdot},\boldsymbol{\cdot})$ indicate the Dynkin labels of the $\SU(4)$ flavor group representation for a given superconformal multiplet. Terms in the sum with negative values for any of the Dynkin label should be excluded. The labels $X\overline{Y}$ indicate the type of superconformal multiplet with the prescribed quantum numbers for the Lorentz spin, conformal dimension, and $\UU(1)$ R-charge. $X$ and $Y$ can either be $L$ for long or $A_{1,2}$ or $B_{1,2}$ indicating which shortening condition applies \cite{Cordova:2016emh}. For $n\geq 2$ the structure of short multiplets appearing in the sums is as follows. The last line of \eqref{N2multiplets} always contains exclusively long multiplets. When $n\equiv 0\mod 4$ the sum on the first line of \eqref{N2multiplets} contributes two short multiplets (plus their conjugate ones), 
\begin{equation}
	\left(n,0,0\right)L\overline{A}_1[2]_{\f n2+2}^{(\f n4)} + \text{c.c.}\,, \qquad \left(n+1,0,1\right)L\overline{A}_2[0]_{\f n2+1}^{(\f n4)} + \text{c.c.} \,,
\end{equation}
while the sum on the second line in \eqref{N2multiplets}  for these values of $n$ contains only long multiplets. When $n\equiv 2\mod 4$ on the other hand, the sum on the first line of in \eqref{N2multiplets} contributes one (complex) short multiplet
\begin{equation}
	\left(n+2,0,0\right)L\overline{B}_1[0]_{\f n2+1}^{(\f n4 +\f12)} + \text{c.c.}\,, 
\end{equation}
and the sum on the second line of \eqref{N2multiplets} contributes another (complex) short multiplet,
\begin{equation}
	\left(n,1,0\right)L\overline{A}_1[1]_{\f n2+\f32}^{(\f n4)} + \text{c.c.}\,.
\end{equation}
All other multiplets not specifically specified above are long. Thus at every even KK level $n$ we get four short multiplets organized as two complex conjugate pairs. 

Summarizing, we find that depending on the spin structure the KK spectrum of 11d supergravity on AdS$_4\times S^7/\bZ_4$ organizes into $\cN=6$ or $\cN=2$ supermultiplets and crucially, for the antiperiodic spin structure the spectrum does not contain additional supersymmetry current multiplets. From this we can conclude that the AdS$_4\times S^7/\bZ_4$ background of 11d supergravity with the antiperiodic spin structure should be dual to a new $\cN=2$ SCFT. In Section~\ref{sec:discussion} we further discuss some of the properties of this SCFT. We now turn our attention to a generalization of the analysis above for the KK spectrum of 11d supergravity on AdS$_4\times S^7/\bZ_k$ for the class of lens spaces presented in Section~\ref{subsec:generalization1}.

\subsection{Generalizations}
\label{subsec:generalization2}

So far, we considered smooth orbifolds of $S^7$ with the orbifold action induced from $\bC^4$ given as in \eqref{orbifoldaction} with $n_{1,2,3}=1$. For general (even) $k>4$, the resulting manifold does not preserve any invariant Killing spinors for the antiperiodic spin structure and hence this generalization does not result in additional new SCFTs. However, as discussed in Section~\ref{subsec:generalization1}, an interesting generalization is found by specifying to $k=4p$ with $p\in \bN$ and the orbifold action as determined in \eqref{first}-\eqref{fourth}. The resulting AdS$_4\times S^7/\bZ_k$ background preserves $\f12$ of the maximal number of Killing spinors for the periodic spin structure while it preserves $\f14$ of the maximal number of Killing spinors for the antiperiodic spin structure. In the dual SCFT, this corresponds to four pairs of SCFTs for every $k=4p$. One member of the pair, corresponding to the periodic spin structure, preserving $\cN=4$ superconformal symmetry, while the other, corresponding to the antiperiodic spin structure, preserves $\cN=2$. 

To provide further evidence for the existence of these SCFTs we can again study the KK spectrum of 11d supergravity and show that it organizes into superconformal multiplets for $\cN=4$ or $\cN=2$ supersymmetry depending on the spin structure. In this section we take the first step in this direction by explicitly working out the branching and reorganization into superconformal multiplets for the first two KK levels. To keep the discussion concrete and the notation as simple as possible we focus our attention on the first orbifold action in \eqref{first} for every choice of $k=4p$. The other three cases in \eqref{second}-\eqref{fourth} can be studied analogously. 

We start by presenting the breaking pattern for the $\SO(8)$ isometry of the round $S^7$ induced by the orbifold action in \eqref{first}
\begin{equation}\label{eq:so8susuu1u1}
	\SO(8) \rightarrow \SU(4)\times \UU(1) \rightarrow \SU(2)_1\times \SU(2)_2 \times \UU(1)_{\SU(4)}\times \UU(1)\,.
\end{equation}
Under this breaking the supercharges decompose as follows,
\begin{equation}\label{eq:genQbranch}
	\begin{aligned}
		\mathbf{8}_s &\rightarrow\, \mathbf{6}_0 \oplus \mathbf{1}_{\f12}\oplus \mathbf{1}_{-\f12} \\
		&\rightarrow \,\left(\mathbf{1},\mathbf{1}\right)_{\f12,0} \oplus \left(\mathbf{1},\mathbf{1}\right)_{-\f12,0} \oplus \left(\mathbf{2},\mathbf{2}\right)_{0,0} \oplus \left(\mathbf{1},\mathbf{1}\right)_{0,\f12} \oplus \left(\mathbf{1},\mathbf{1}\right)_{0,-\f12}\,.
	\end{aligned} 
\end{equation}
When we consider the periodic spin structure the dual SCFT preserves $\cN=4$ supersymmetry and hence the $\SU(2)_1\times \SU(2)_2$ symmetry represents the R-symmetry while the $\UU(1)_{\SU(4)}\times \UU(1)$ is a flavor symmetry. We can now formulate a generalization of the $\bZ_4$ rule introduced above for the orbifold induced by the action \eqref{first} 

\noindent\textbf{A $\bZ_{4p}$ rule:} \textit{For both spin structures, the charge under $\UU(1)_{\SU(4)}$ for both the bosonic and fermionic modes invariant under the orbifold action has to be a multiple of $p/2$. For the periodic spin structure all bosonic and fermionic KK modes invariant under the orbifold action have integer $\UU(1)$ charge. For the antiperiodic spin structure instead, the bosonic modes should have integer $\UU(1)$ charge, while the $\UU(1)$ charges of the fermionic modes should be half integer.} 
	
Equipped with this rule we are now left with a group theory exercise to organize the KK modes into $\cN=4$ and $\cN=2$ superconformal multiplets. We can immediately note that this selection rule selects exactly the preserved supercharges transforming in the $\left(\mathbf{2},\mathbf{2}\right)_{0,0}$ representation in \eqref{eq:genQbranch} for the periodic spin structure while for the antiperiodic spin structure only the last two supercharges in \eqref{eq:genQbranch}, $\left(\mathbf{1},\mathbf{1}\right)_{0,\pm\f12}$ are preserved. This is precisely the expected structure for an SCFT with $\mathcal{N}=4$ or $\mathcal{N}=2$ superconformal symmetry, respectively.

Let us start with the periodic spin structure. At level $n=0$, the part of the spectrum invariant under the orbifold action simply consists of those multiplets neutral under $\UU(1)_{\SU(4)}\times \UU(1)$. The invariant part of the KK spectrum is given in Table~\ref{tab:KKGenn0} from which we can build the corresponding $\cN=4$ multiplets. They are given by $A_2[0]_1^{(0,0)}$, $B_1[0]_1^{(0,2)}$, and $B_1[0]_1^{(2,0)}$, i.e. the $\cN=4$ stress tensor multiplet together with two $\UU(1)$ flavor multiplets. In the special case when $k=4$, i.e. $p=1$ the orbifold action reduces to the one discussed in Section~\ref{subsec:periodZ4}. Indeed, we find that in this case an additional pair of SUSY current multiplets $B_1[0]_1^{\left(0,0\right)}$ appears, in line with the fact that for $k=4$ and the periodic spin structure we have $\cN=6$ supersymmetry.

\begin{table}[!htb]
	\centering
	\begin{tabular}{c|c|c|c}
		Field & $\SU(2)_1\times \SU(2)_2 \times \UU(1)_{\SU(4)}\times \UU(1)$ irrep & $j$ &$\Delta$ \\
		\hline
		$e^{(0)}_\mu{}^a$ & $\left(\mathbf{1},\mathbf{1}\right)_{0,0}$ & 4 & $3$\\
		$\psi^{(0)}_\mu$ & $\left(\mathbf{2},\mathbf{2}\right)_{0,0}$ & 3 & $\f52$\\
		$A^{(0)}_\mu$ & $2\left(\mathbf{1},\mathbf{1}\right)_{0,0} \oplus\left(\mathbf{1},\mathbf{3}\right)_{0,0} \oplus\left(\mathbf{3},\mathbf{1}\right)_{0,0} $& 2 & $2$ \\
		$\chi^{(0)}$ & $3\left(\mathbf{2},\mathbf{2}\right)_{0,0}$ & 1 & $\f32$\\
		$S^{(0)}$ & $\left(\mathbf{1},\mathbf{1}\right)_{0,0} \oplus\left(\mathbf{1},\mathbf{3}\right)_{0,0} \oplus\left(\mathbf{3},\mathbf{1}\right)_{0,0}$& 0 & $1$\\
		$P^{(0)}$ & $\left(\mathbf{1},\mathbf{1}\right)_{0,0} \oplus\left(\mathbf{1},\mathbf{3}\right)_{0,0} \oplus\left(\mathbf{3},\mathbf{1}\right)_{0,0}$& 0 & $2$\\
	\end{tabular}
	\caption{The KK spectrum for the level $n=0$ of 11d supergravity on AdS$_4\times S^7/\bZ_{4p}$ with the orbifold action in \eqref{first} and the periodic spin structure.}
	\label{tab:KKGenn0}
\end{table}

When we consider the antiperiodic spin structure the spectrum should organize into $\cN=2$ multiplets. In this case we should have an $\SU(2)_1\times \SU(2)_2\times\UU(1)_{\SU(4)}$ flavor symmetry while the second $\UU(1)$ in \eqref{eq:so8susuu1u1} represents the $\cN=2$ R-symmetry. To obtain the invariant KK modes under the orbifold action we keep the same bosonic modes as for the periodic spin structure above but but we change the fermionic spectrum by keeping only those representations for which the second $\UU(1)$ charge is a half integer. The result of this selection procedure for the lowest KK level $n=0$ is given in Table~\ref{tab:KKGenaltn0}. We find that indeed the KK modes can be organized into $\mathcal{N}=2$ superconformal multiplets where all constituents of a multiplet transform in the same way under the $\SU(2)\times \SU(2)\times \UU(1)_{\SU(4)}$ flavor symmetry. The resulting spectrum is given by the $\cN=2$ stress tensor multiplet $A_1\overline{A}_1[2]_2^{0}$, transforming in the $(\mathbf{1},\mathbf{1})_0$ representation of the flavor symmetry, accompanied by a set of flavor current multiplets $A_1\overline{A}_1[0]_1^{0}$ transforming in the adjoint representation of the flavor symmetry, i.e. $(\mathbf{1},\mathbf{1})_0\oplus (\mathbf{1},\mathbf{3})_0 \oplus (\mathbf{3},\mathbf{1})_0$.  

\begin{table}[!htb]
	\centering
	\begin{tabular}{c|c|c|c}
		Field & $\SU(2)_1\times \SU(2)_2 \times \UU(1)_{\SU(4)}\times \UU(1)$ irrep & $j$ &$\Delta$ \\
		\hline
		$e^{(0)}_\mu{}^a$ & $\left(\mathbf{1},\mathbf{1}\right)_{0,0}$ & 4 & $3$\\
		$\psi^{(0)}_\mu$ & $\left(\mathbf{1},\mathbf{1}\right)_{0,\pm\f12}$ & 3 & $\f52$\\
		$A^{(0)}_\mu$ & $2\left(\mathbf{1},\mathbf{1}\right)_{0,0} \oplus\left(\mathbf{1},\mathbf{3}\right)_{0,0} \oplus\left(\mathbf{3},\mathbf{1}\right)_{0,0} $& 2 & $2$ \\
		$\chi^{(0)}$ & $\left(\mathbf{1},\mathbf{1}\right)_{0,\pm\f12}\oplus \left(\mathbf{1},\mathbf{3}\right)_{0,\pm\f12} \oplus \left(\mathbf{3},\mathbf{1}\right)_{0,\pm\f12}$ & 1 & $\f32$\\
		$S^{(0)}$ & $\left(\mathbf{1},\mathbf{1}\right)_{0,0} \oplus\left(\mathbf{1},\mathbf{3}\right)_{0,0} \oplus\left(\mathbf{3},\mathbf{1}\right)_{0,0}$& 0 & $1$\\
		$P^{(0)}$ & $\left(\mathbf{1},\mathbf{1}\right)_{0,0} \oplus\left(\mathbf{1},\mathbf{3}\right)_{0,0} \oplus\left(\mathbf{3},\mathbf{1}\right)_{0,0}$& 0 & $2$\\
	\end{tabular}
	\caption{The KK spectrum for the level $n=0$ of 11d supergravity on AdS$_4\times S^7/\bZ_{4p}$ with the orbifold action in \eqref{first} and the antiperiodic spin structure.}
	\label{tab:KKGenaltn0}
\end{table}

One can repeat the same exercise for the higher KK levels and again find that the spectrum nicely organizes into either $\cN=4$ or $\cN=2$ multiplets depending on the spin structure. For level $n=2$ the result is presented explicitly in Appendix~\ref{app:moretables}. In this case we find that the spectrum contains both short and long superconformal multiplets.

\section{Discussion}
\label{sec:discussion}

The discussion above points to the existence of two different SCFTs that are holographically dual to the $\AdS_4\times S^7/\bZ_4$ solution of 11d supergravity. One of these, corresponding to the periodic spin structure is the $\mathcal{N}=6$ $\UU(N)\times\UU(N)$ ABJM SCFT at level $k=4$.\footnote{The holographic description for the $\mathcal{N}=6$ ABJM SCFTs with gauge groups $\SU(N)_k\times\SU(N)_{-k}$, $(\UU(N)_{k}\times\UU(N)_{-k})/\bZ_N$, and $(\UU(N)_{k}\times\UU(N)_{-k})/\bZ_k$ is based on the same supergravity solution with modified boundary conditions and was discussed in \cite{Bergman:2020ifi}.} The SCFT corresponding to the antiperiodic spin structure should preserve $\mathcal{N}=2$ supersymmetry and to the best of our knowledge has not been studied in the literature. While we cannot offer a complete description of this $\mathcal{N}=2$ SCFT here we discuss some of its properties that are suggested by the supergravity dual.

The natural way to mimic the supergravity construction in the dual 3d SCFT is to start with the ABJM theory at level $k=1$ with $\mathcal{N}=8$ supersymmetry and $\SO(8)$ R-symmetry. When written in $\mathcal{N}=2$ superspace language the theory has a manifest $\SU(4)\times \UU(1)_b$ symmetry and we can ``mod out'' by a $\bZ_k$ subgroup of $\UU(1)_b$  to arrive at the a theory with $\mathcal{N}=6$ supersymmetry, $\SU(4)$ R-symmetry, and a $\UU(1)_b$ flavor symmetry.  This process can be viewed as a gauging of the $\bZ_k$ symmetry which leads to preserving only gauge invariant operators of the $k=1$ ABJM theory invariant  under the $\bZ_k$ action. This gauging of a discrete subgroup of the flavor group leads to a $\bZ_k$ 1-form symmetry, see \cite{Gaiotto:2014kfa,Tachikawa:2017gyf}, which is the 1-form symmetry for the $\UU(N)_k\times\UU(N)_{-k}$  ABJM theory \cite{Bergman:2020ifi}. The procedure we describe is somewhat formal and applies to the ``standard'', i.e. periodic, choice of spin structure on $S^7/\bZ_k$. For $k=4$ the supergravity discussion above suggests that there is a different way to gauge the $\bZ_4$ which essentially amounts to implementing the ``$\bZ_4$ rule'' we used in Section~\ref{sec:KK} to deduce the 11d supergravity KK spectrum. This is a simple procedure to implement if the full solution of the $k=1$ $\mathcal{N}=8$ ABJM SCFT is under control, however we are not sure how to justify it at the level of the Lagrangian of the ABJM theory. In particular it is not clear to us how these two different ways to perform the $\bZ_4$  gauging will affect the spectrum of unprotected operators, structure constants, and non-local operators in the theory.

In the absence of a more direct field theory description of the $k=4$ $\mathcal{N}=2$ SCFT corresponding to the antiperiodic spin structure we make a few observations based on its supergravity dual. First we note that the all SCFT operators dual to the bosonic KK modes of 11d supergravity have the same spectrum as for the bosonic operators in the $k=4$ $\mathcal{N}=6$ ABJM theory. The fermionic spectrum however is different. This suggests that the supersymmetric partition functions of the two theories are also distinct in general. For instance the $S^1\times S^2$ superconformal index that counts certain BPS operators should be different for the two theories and it will be very interesting to compute this index along the lines of \cite{Kim:2009wb}. An interesting consequence of the different spectrum is that the $\mathcal{N}=2$ and $\mathcal{N}=6$ SCFTs have different dimensions of their conformal manifolds. The $L\overline{B}_1 [0]_2^{(2)}$ and $B_1\overline{L} [0]_2^{(-2)}$ supermultiplets in Table~\ref{tab:n2altmultipletapp} are in the $\mathbf{35}$ and $\overline{\mathbf{35}}$ of the $\SU(4)$ flavor symmetry and contain scalar operators with $\Delta=3$ that are candidate marginal operators. Taking into account the $\SU(4)$ flavor symmetry and using the results of \cite{Green:2010da} we conclude that there are 20 (complex) exactly marginal operators in this theory. This is in contrast to the $k=4$ $\mathcal{N}=6$ ABJM theory which, as summarized recently in \cite{Bobev:2021gza}, has only 3 exactly marginal operators. Other interesting observable in the two theories are the $S^3$ free energy and the topologically twisted index. To leading order in the large $N$ limit both of these quantities can be compute by supersymmetric localization for the $k=4$ $\mathcal{N}=6$ ABJM theory and they agree with the two-derivative supergravity result, see \cite{Fuji:2011km,Marino:2011eh} and \cite{Benini:2015eyy,Azzurli:2017kxo}. Since the supergravity calculation of these observables depends only on the volume of $S^7/\bZ_k$ we conclude that both for the ABJM theory at $k=4$ and for the new $\mathcal{N}=2$ SCFT we should have the same expressions for the topologically twisted index and the $S^3$ free energy to leading order at large $N$
\begin{equation}
F_{S^3} = \frac{2\pi\sqrt{2}}{3}N^{3/2}\,, \qquad\qquad \log Z_{S^1\times \Sigma_{\mathfrak{g}}} = \frac{2\pi\sqrt{2}}{3}(\mathfrak{g}-1)N^{3/2}\,.
\end{equation}
It will be very interesting to understand how to derive this results directly from the field theory for the new $k=4$ $\mathcal{N}=2$ SCFT.\footnote{We note in passing that for $k=4$ one of the two $N^{1/2}$ corrections to the squashed $S^3$ free energy of the $\mathcal{N}=6$ ABJM theory vanishes. This corresponds to a vanishing contribution from one of the higher derivative corrections to 4d minimal supergravity \cite{Bobev:2020egg,Bobev:2021oku}. We do not know whether this curious fact has any relation to the existence of two different spin structure compatible with supersymmetry for $k=4$.} The supergravity AdS$_4\times S^7/\bZ_k$ solution can also be used to deduce the one-loop contribution to the $S^3$ free energy along the lines of \cite{Liu:2016dau}, and it will be interesting to perform this calculation for the antiperiodic spin structure.

A curious feature of the new $k=4$ $\mathcal{N}=2$ SCFT is that, at least at large $N$, the spectrum of local operators contains many long superconformal multiplets with operators of integer and half-integer conformal dimensions.\footnote{For the $\mathcal{N}=6$ ABJM theory it is expected that there are no such accidents.} It will be very interesting to understand what are the $1/N$ corrections to these operator dimensions. 

It is well-known that there is an interesting generalization of the ABJM theory an  $\UU(M)_k\times\UU(N)_{-k}$ $\mathcal{N}=6$ SCFT \cite{Aharony:2008gk}. These theories correspond to $|M-N|$ fractional M2-branes fixed at the $\mathbf{C}^4/\bZ_k$ singularity. This construction leads to a unitary $\mathcal{N}=6$ SCFT for $|M-N|\leq k$, where the integer $|M-N|$ corresponds to the ``torsion flux'' in M-theory. It will be interesting to understand whether for $k=4$ there is an $\mathcal{N}=2$ ``cousin'' of this ABJ theory that corresponds to the antiperiodic spin structure. 

Another well-known simple way to modify a Freund-Rubin AdS$_4\times \mathcal{M}_7$ solution of 11d supergravity is to reverse the orientation of the internal manifold $\mathcal{M}_7$. These ``skew-whiffed'' AdS$_4$ solutions are constructed by changing the orientation of the internal space which in turn changes the sign of the four-form flux $G_4$, as well as the chirality of the Killing spinors on $\cM_7$ \cite{Duff:1984sv}, see also \cite{Berkooz:1998qp,Gauntlett:2009bh} for a holographic discussion. For the lens spaces $S^7/\bZ_k$ of interest here when we choose the periodic structure and general values of $k$ all invariant spinors have positive chirality \cite{Figueroa-OFarrill:2005vxy}. We thus conclude that ``skew-whiffed'' AdS$_4$ solutions for these lens spaces break all supersymmetry. On the round $S^7$ both the ``skew-whiffed'' and the standard solutions preserve the maximal supersymmetry. For $k=2$, the manifold $\bR\bP^7$ is special in the sense that it preserves the maximal amount of supersymmetry for both spin structures. Moreover, the invariant Killing spinors have opposite chirality for the two spin structures. Therefore both the standard and the ``skew-whiffed'' Freund Rubin solutions preserve the maximal number of invariant Killing spinors, however, these spinors live in different spin bundles. For $k=4$, however the spinors for both spin structures have the same chirality and hence in this case skew-whiffing again breaks all supersymmetry. Similarly for the generalizations we discussed in Section~\ref{subsec:generalization1}, the spinors in both spin bundles have the same chirality and therefore only one orientation of the internal manifold preserves supersymmetry.

In this work we discussed smooth orbifolds $S^7/G$ where $G$ is restricted to be a cyclic group. However, there is a richer story to be explored for more general finite subgroups $G\subset\SO(8)$. In \cite{deMedeiros:2009pp} such orbifolds preserving at least $\cN=4$ supersymmetry were discussed in detail. This could be a starting point for further exploration in the context of AdS/CFT. However, the question about the influence of the choice of spin structure on the supersymmetry of the dual field theory goes far beyond orbifolds of spheres and it could be studied for any internal spin manifold with $H^1(\cM,\bZ_2)\neq 0$. This applies for internal manifolds of any dimension and for various choices of the type of supergravity theory. It would be very interesting to explore this subject further in the context of holography and map out the full range of possibilities and the implications for the dual CFTs.  We end our discussion by pointing out that the choice of spin structure is also important when studying non-trivial solutions of a supergravity theory that break the isometries of AdS. This was emphasized in \cite{Martelli:2012sz} where fillings of Euclidean $\AdS_4$ with different topology were discussed and it was found that not all of them can be uplifted to 11d supergravity for all choices of spin structure and orbifold action on $S^7$. This further highlights the importance of the choice of spin structures in holography in setups with broken conformal invariance.
\bigskip
\bigskip
\leftline{\bf Acknowledgments}
\smallskip

\noindent We would like to thank Davide Cassani, Anthony Charles, Kiril Hristov, Valentin Reys, and Chiara Toldo for useful discussions. We are especially grateful to Silviu Pufu for the collaboration in the early stages of this project and for a number of fruitful discussions. The work of NB is supported in by an Odysseus grant G0F9516N from the FWO and by the KU Leuven C1 grant ZKD1118 C16/16/005. PB is supported by the STARS-StG grant THEsPIAN, a `Francqui' Fellowship of the Belgian American Educational Foundation and a Fulbright Fellowship.


\appendix

\section{Conventions}
\label{app:group}

In this work we interchangeably denote representations by their dimension or Dynkin label. In order to prevent confusion here we present our notation and conventions for the dictionary to translate between the two notations.

The Dynkin label for a finite irreducible representation of a group $G$ of rank $r_G$ is given by a set of $r_G$ non-negative integers $\alpha_i$ such that the highest weight vector $\alpha$ can be written in terms of the fundamental weights $\omega_i$ as
\begin{equation}
	\alpha = \sum_{i=1}^{r_G} \alpha_i\,\omega_i\,.
\end{equation}
In this work we encounter representations of $\SO(8)$, $\SU(4)$, and $\SU(2)$ which have rank $r_{\SO(8)}=4$ , $r_{\SU(4)}=3$ and $r_{\SU(2)}=1$.\footnote{By the well-known exceptional isomorphism we have $\SO(6)\simeq \SU(4)$. In our conventions, the $\SO(6)$ and $\SU(4)$ Dynkin labels are related by a permutation of the first two entries.} In our conventions the triality automorphism of $\SO(8)$ permutes the Dynkin labels $\alpha_1$, $\alpha_3$ and $\alpha_4$ and the three eight-dimensional representations are given by
\begin{equation}
	\mathbf{8}_s = (0,0,0,1)\,,\qquad  \mathbf{8}_v = (1,0,0,0)\,,\qquad  \mathbf{8}_c = (0,0,1,0)\,. 
\end{equation}
To relate the Dynkin label to the dimension of the representation one can use the Weyl dimension formula which for an $\SO(8)$ representation $\lambda=(\alpha_1,\alpha_2,\alpha_3,\alpha_4)$, $\SU(4)$ representation $\mu=(\alpha_1,\alpha_2,\alpha_3)$, and $\SU(2)$ representation $\nu=(\alpha_1)$ is given by
\begin{align}
	\dim(\lambda) =& \f{1}{4320}(1+\alpha_1)(1+\alpha_2)(1+\alpha_3)(1+\alpha_4)(2+\alpha_1+\alpha_2)(2+\alpha_2+\alpha_3)(2+\alpha_2+\alpha_4)\times\nn\\
	&\times(3+\alpha_1+\alpha_2+\alpha_3)(3+\alpha_1+\alpha_2+\alpha_4)(3+\alpha_2+\alpha_3+\alpha_4)\times\\
	&\times(4+\alpha_1+\alpha_2+\alpha_3+\alpha_4)(5+\alpha_1+2\alpha_2+\alpha_3+\alpha_4)\,,\nn\\
	\dim(\mu) =& \f{1}{12}(1+\alpha_1)(1+\alpha_2)(1+\alpha_3)\times\\
	&\times(2+\alpha_1+\alpha_2)(2+\alpha_2+\alpha_3)(3+\alpha_1+\alpha_2+\alpha_3)\,,\nn\\
	\dim(\nu) =& \alpha_1+1\,.
\end{align}

To denote superconformal multiplets we use the notation of \cite{Cordova:2016emh} and refer to that paper for more details on the various shortening conditions. The notation is as follows, $X[j]_\Delta^{(r)}$, where $X$ specifies the shortening condition, $j$ is the Lorentz $\SU(2)$ Dynkin label,\footnote{Note that the $\SU(2)$ Dynkin label equals twice the spin of the top component of the multiplet.} $\Delta$ is the conformal dimension, and $(r)$ denotes the Dynkin label for the R-symmetry. All quantum numbers and representations refer to the top component of the multiplet. Note however that for the $\cN=2$ multiplets we normalize the R-symmetry with a factor of $\f12$ with respect to \cite{Cordova:2016emh}, i.e. for us the $\mathcal{N}=2$ supercharges are denoted as $Q\in [1]_{\f12}^{\left(-\f12\right)}$ instead of $[1]_{\f12}^{(-1)}$.

\section{Branching rules}
\label{app:branching}

In this appendix we consider the branching rules for $\SO(8)\rightarrow \SU(4)\times \UU(1)$. As discussed in the main text we normalize the $\UU(1)$ charge such that,
\begin{equation}
	\mathbf{8}_s \rightarrow \mathbf{6}_0 + \mathbf{1}_{-\f12}+ \mathbf{1}_{\f12}\,.
\end{equation}
The KK spectrum of 11d supergravity on $S^7$ was originally obtained in \cite{Duff:1986hr,Biran:1983iy} and the branching rules for the relevant $\SO(8)$ representations were derived in \cite{Nilsson:1984bj,Liu:2016dau}. The resulting $\SU(4)\times \UU(1)$ representations are given in Table \ref{tab:branching}, where we use the shortened notation,
\begin{equation}\label{branchnotation}
	[a,b,c;d] = \sum_{l=0}^N (n-l+a,b,l+c)_{\f{n-2l+d}{4}}\,.
\end{equation}
In this equation, $(n-l+a,b,l+c)$ is the Dynkin label for an irreducible $\SU(4)$ representations and $\f{n-2l+d}{4}$ is the corresponding $\UU(1)$ charge. The integer $N$ is defined as the highest entry of the relevant $\SO(8)$ Dynkin label.

\begin{table}[!htb]
	\centering
	\begin{tabular}{c|c|p{8cm}|c}
		Spin & $\SO(8)$ irrep & $\SU(4)\times \UU(1)$ irrep & $\Delta$\\
		\hline
		$2^{+}$ & $(n,0,0,0), n\geq 0 $  & $[0,0,0;0]$ & $\frac{n}{2}+3$ \\[2pt] 
		\hline
		$\frac{3}{2}^{(1)}$ & $(n,0,0,1)$, $n\geq 0$ & $[0,1,0;0]+[0,0,0;-2]+[0,0,0;2]$ & $\frac{n}{2}+\frac{5}{2}$ \\[2pt] 
		\hline
		$\frac{3}{2}^{(2)}$ & $(n-1,0,1,0),n\geq 1$ & $[0,0,0;-2]+[-1,1,-1;0]+[-1,0,1;0] $& $\frac{n}{2}+\frac{7}{2}$\\[2pt]    
		\hline
		$1^{-(1)}$& $(n,1,0,0),n\geq 0$& $[0,0,0;0]+[1,0,1;0]+[0,1,0;-2]+[0,1,0;2]$&$\frac{n}{2}+2$\\[2pt] 
		\hline
		$1^{+}$&$(n-1,0,1,1),n\geq1$&$[0,0,0;0]+[-1,0,1;-2]+[0,1,0;-2]$\par${}+[-1,1,1;0]+[-1,1,-1;-2]+[-2,1,0;0]$\par${}+[-1,2,-1;0]+[0,0,0;-4]+[-1,0,1;2]$&$\frac{n}{2}+3$\\ 
		\hline
		$1^{-(2)}$&$(n-2,1,0,0),n\geq2$&$[-2,0,0;-2]+[-1,0,1;-2]+[-2,1,0;-4]$\par${}+[-2,1,0;0]$&$\frac{n}{2}+4$\\ 
		\hline
		$\frac{1}{2}^{(1)}$&$(n+1,0,1,0),n\geq 0$&$[2,0,0;0]+[1,1,-1;2]+[1,0,1;2]$&$\frac{n}{2}+\frac{3}{2}$\\[2pt] 
		\hline
		$\frac{1}{2}^{(2)}$&$(n-1,1,1,0),n\geq 1$&$[0,0,0;-2]+[-1,1,-1;0]+[-1,0,1;0]$\par${}+[1,0,1;-2]+[0,0,2;0]+[0,1,0;0]$\par${}+[-1,1,1;-2]+[0,1,0;-4]+[-1,1,1;2]$\par${}+[-2,2,0;-4]+[-1,2,-1;2]$&$\frac{n}{2}+\frac{5}{2}$\\ 
		\hline
		$\frac{1}{2}^{(2)}$&$(n-2,1,0,1),n\geq 2$&$[-1,0,1;0]+[-1,0,1;-4]+[-2,0,0;-4]$\par${}+[-2,0,0;0]+[-1,1,1;-2]+[-2,1,0;-2]$\par${}+[-2,1,0;-2]+[-2,1,0;2]+[-2,1,0;-6]$\par${}+[-2,2,0;0]+[-2,2,0;-4]$&$\frac{n}{2}+\frac{7}{2}$\\ 
		\hline
		$\frac{1}{2}^{(4)}$&$(n-2,0,0,1),n\geq2$&$[-2,1,0;-2]+[-2,0,0;-4]+[-2,0,0;0]$&$\frac{n}{2}+\frac{9}{2}$\\[2pt] 
		\hline
		$0^{+(1)}$&$(n+2,0,0,0),n\geq0$&$[2,0,0;2]$&$\frac{n}{2}+1$\\[2pt] 
		\hline
		$0^{-(1)}$&$(n,0,2,0),n\geq0$&$[1,0,1;0]+[2,0,0;-2]+[0,0,2;2]$\par${}+[-1,2,-1;0]+[1,1,-1;0]+[-1,1,1;0]$&$\frac{n}{2}+2$ \\ 
		\hline
		$0^{+(2)}$&$(n-2,2,0,0),n\geq2$&$[-1,1,1;0]+[-1,1,1;-4]+[-2,2,0;-6]$\par${}+[-2,-2,0;2]+[-1,0,1;-2]+[-2,2,0;-2]$\par${}+[-2,1,0;-4]+[-2,1,0;0]+[-2,0,0;-2]$\par${}+[0,0,2;-2]$&$\frac{n}{2}+3$\\ 
		\hline
		$0^{-(2)}$&$(n-2,0,0,2),n\geq2$&$[-2,0,0;-2]+[-2,0,0;-6]+[-2,0,0;2]$\par${}+[-2,2,0;-2]+[-2,1,0;-4]+[-2,1,0;0]$&$\frac{n}{2}+4$\\ 
		\hline
		$0^{+(3)}$&$(n-2,0,0,0),n\geq2$&$[-2,0,0;-2]$&$\frac{n}{2}+5$\\[2pt] 
		\hline
	\end{tabular}
	\caption{Branching rules for $\SO(8)\rightarrow \SU(4)\times \UU(1)$ for the representations of interest in this paper.}
	\label{tab:branching}
\end{table}

To determine the spectrum on $S^7/\bZ_k$ we have to select those representations that remain invariant under the $\bZ_k$ orbifolds introduced in the main text. For the bosonic modes this simply amounts to selecting those representations with $\UU(1)$ charges a multiple of $k/4$. For the fermionic modes on the other hand, the selection rules depend crucially on the spin structure. In particular, for the periodic spin structure we keep the same selection rule as for the bosons while for the antiperiodic spin structure we retain those fermions with $\UU(1)$ charges $q=\f{1}{2k}+\f{mk}{4}$. However, as explained in the main text, only for level $k=2$ or $k=4$ does this result in a set of two supersymmetric spectra.
For $k=4$, the resulting spectrum is explicitly given for level $n=0$ in Table~\ref{tab:KKZ4n0} and Table~\ref{tab:KKZ4altn0} while for the next few even levels the result is given Appendix~\ref{app:moretables}.  

Specifying to $k=4$, we can make the following observation for the odd levels. Inspecting all the representations that appear in Table~\ref{tab:branching}, we see that all of them have $d$ even, in the notation introduced in~\eqref{branchnotation}. For this reason it immediately follows that for any even KK level, the $\UU(1)$ charge can only be integer or half integer. For odd levels on the other hand, the $\UU(1)$ charges are always of the form $q=\f14 + \f{p}{2}$ where $p\in \bZ$ and hence they are never integer or half integer. We therefore conclude that the selection rule for both spin structures will never retain any representations from the odd KK levels and hence they play no role in most of this work.

\FloatBarrier

\section{More examples}
\label{app:moretables}

In the main text we presented a rule to select the invariant part of the KK spectrum under a smooth orbifold action. For both the $\bZ_4$ as well as its generalizations to $\bZ_{4p}$ we illustrated how the resulting spectrum organizes into superconformal multiplets for low levels. In particular, in the main text we explicitly gave the spectrum for KK level $n=0$ as well as a general formula for the $\bZ_4$ case. In this appendix we collect some additional examples to further illustrate this reorganization into superconformal multiplets. 

\subsection{Higher levels for the $\bZ_4$ orbifold}

Analogous to the examples in the main text we can continue to higher levels and using the $\bZ_4$ rule explicitly write down the KK spectrum for both spin structures on $S^7/\bZ_4$. To illustrate this in some more detail we present the KK spectrum at level $n=2$ for $S^7/\bZ_k$ in Tables~\ref{tab:KKZ4n2} and Table~\ref{tab:KKZ4altn2} for the periodic and antiperiodic spin structure respectively.

\begin{table}[!htb]
	\centering
	\begin{tabular}{c|c|c}
		Field & $\SU(4)\times \UU(1)$ irreps & $\Delta$ \\
		\hline
		$e^{(2)}_\mu{}^a$ & $\mathbf{15}_0$ & $4$\\
		\hline
		\multirow{2}{*}{$\psi^{(2)}_\mu$} &  $ \mathbf{64}_0+\mathbf{10}_1 + \mathbf{10}_0 + \overline{\mathbf{10}}_0+ \overline{\mathbf{10}}_{-1}$ & $\f72$\Tstrut \\
		& $\mathbf{10}_0 + \overline{\mathbf{10}}_0 + \mathbf{6}_0$ & $\f92$\Tstrut\\
		\hline
		\multirow{3}{*}{$A^{(2)}_\mu$} &  $\mathbf{84}_0+\mathbf{45}_1+ \mathbf{45}_0 + \overline{\mathbf{45}}_0+ \overline{\mathbf{45}}_{-1} + \mathbf{15}_0 $ & $3$\Tstrut\\
		& $\mathbf{45}_0+\overline{\mathbf{45}}_0 + \mathbf{20}^\prime_0 + \mathbf{15}_{1} + 2\cdot\mathbf{15}_0 + \mathbf{15}_{-1}+\mathbf{1}_0 $ & $4$\Tstrut\\
		& $\mathbf{15}_0+\mathbf{1}_0$ & $5$\Tstrut\\
		\hline
		\multirow{4}{*}{$\chi^{(2)}$} &  $ \mathbf{70}_1+\mathbf{70}_0+\overline{\mathbf{70}}_0+\overline{\mathbf{70}}_{-1} + \mathbf{64}_0$ & $\f52$\Tstrut\\
		& $\mathbf{70}_0+\overline{\mathbf{70}}_0+\mathbf{64}_1+2\cdot\mathbf{64}_0+\mathbf{64}_{-1}+\mathbf{10}_0+\overline{\mathbf{10}}_0+\mathbf{6}_0$ & $\f72$\Tstrut\\
		& $\mathbf{64}_0+2\cdot\mathbf{6}_0+\mathbf{6}_1+\mathbf{6}_{-1}$ & $\f92$\Tstrut\\
		& $\mathbf{6}_0$& $\f{11}{2}$\Tstrut\\
		\hline
		\multirow{3}{*}{$S^{(2)}$} &  $ \mathbf{35}_{1}+\mathbf{84}_0+\overline{\mathbf{35}}_{-1}$& $2$\Tstrut\\
		& $\mathbf{84}_0+\mathbf{20}^\prime_1+\mathbf{20}^\prime_0+\mathbf{20}^\prime_{-1}+\mathbf{15}_0+\mathbf{1}_0$& $4$\Tstrut\\
		& $\mathbf{1}_0$& $6$\Tstrut\\
		\hline
		\multirow{2}{*}{$P^{(2)}$} &  $\mathbf{35}_0+\mathbf{84}_1+\mathbf{84}_0+\mathbf{84}_{-1}+\mathbf{45}_0+\overline{\mathbf{45}}_0+\overline{\mathbf{35}}_0+\mathbf{20}^\prime_0$& $3$\Tstrut\\
		& $\mathbf{20}^\prime_0+\mathbf{1}_{-1}+\mathbf{1}_{0}+\mathbf{1}_{1}$& $5$\Tstrut\\
	\end{tabular}
	\caption{The KK spectrum for level $n=2$ of 11d supergravity on AdS$_4\times S^7/\bZ_4$ with the periodic spin structure.}
	\label{tab:KKZ4n2}
\end{table}
\begin{table}[!htb]
	\centering
	\begin{tabular}{c|c|c}
		Field & $\SU(4)\times \UU(1)$ irreps & $\Delta$ \\
		\hline
		$e^{(2)}_\mu{}^a$ & $\mathbf{15}_0$ & $4$\Tstrut \\
		\hline
		\multirow{2}{*}{$\psi^{(2)}_\mu$} &  $ \mathbf{45}_{\f12} + \overline{\mathbf{45}}_{-\f12} + \mathbf{15}_{\f12} + \mathbf{15}_{-\f12}$ & $\f72$\Tstrut \\
		& $\mathbf{15}_{-\f12} + \mathbf{15}_{\f12}$ & $\f92$\Tstrut\\
		\hline
		\multirow{3}{*}{$A^{(2)}_\mu$} &  $\mathbf{84}_0+\mathbf{45}_1+ \mathbf{45}_0 + \overline{\mathbf{45}}_0+ \overline{\mathbf{45}}_{-1} + \mathbf{15}_0 $ & $3$\Tstrut\\
		& $\mathbf{45}_0+\overline{\mathbf{45}}_0 + \mathbf{20}^\prime_0 + \mathbf{15}_{1} + 2\cdot\mathbf{15}_0 + \mathbf{15}_{-1} $ & $4$\Tstrut\\
		& $\mathbf{15}_0+\mathbf{1}_0$ & $5$\Tstrut\\
		\hline
		\multirow{4}{*}{$\chi^{(2)}$} &  $ \mathbf{35}_{\f12} + \mathbf{84}_{\f12} + \mathbf{84}_{-\f12} + \mathbf{45}_{\f12} + \overline{\mathbf{35}}_{-\f12} + \overline{\mathbf{45}}_{-\f12}$ & $\f52$\Tstrut\\
		& $\mathbf{84}_{\f12} + \mathbf{84}_{-\f12} + \mathbf{45}_{\f12} + \mathbf{45}_{-\f12} + \overline{\mathbf{45}}_{\f12} + \overline{\mathbf{45}}_{-\f12} + \mathbf{20}^\prime_{\f12} + \mathbf{20}^\prime_{-\f12} + \mathbf{15}_{-\f12} + \mathbf{15}_{-\f12}$ & $\f72$\Tstrut\\
		& $\mathbf{20}^\prime_{\f12} + \mathbf{20}^\prime_{-\f12} + \mathbf{15}_{\f12} + \mathbf{15}_{-\f12} + \mathbf{1}_{\f12} + \mathbf{1}_{-\f12}$ & $\f92$\Tstrut\\
		& $\mathbf{1}_{\f12} + \mathbf{1}_{-\f12}$& $\f{11}{2}$\Tstrut\\
		\hline
		\multirow{3}{*}{$S^{(2)}$} &  $ \mathbf{35}_{1} + \mathbf{84}_0 + \overline{\mathbf{35}}_{-1}$& $2$\Tstrut\\
		& $\mathbf{84}_0 + \mathbf{20}^\prime_1 + \mathbf{20}^\prime_0 + \mathbf{20}^\prime_{-1} + \mathbf{15}_0 + \mathbf{1}_0$& $4$\Tstrut\\
		& $\mathbf{1}_0$& $6$\Tstrut\\
		\hline
		\multirow{2}{*}{$P^{(2)}$} &  $\mathbf{35}_0 + \mathbf{84}_1 + \mathbf{84}_0 + \mathbf{84}_{-1} + \mathbf{45}_0 + \overline{\mathbf{45}}_0 + \overline{\mathbf{35}}_0 + \mathbf{20}^\prime_0$& $3$\Tstrut\\
		& $\mathbf{20}^\prime_0 + \mathbf{1}_{-1} + \mathbf{1}_{0} + \mathbf{1}_{1}$& $5$\Tstrut\\
	\end{tabular}
	\caption{The KK spectrum for level $n=2$ of 11d supergravity on AdS$_4\times S^7/\bZ_4$ with the antiperiodic spin structure.}
	\label{tab:KKZ4altn2}
\end{table}

\FloatBarrier

After having found the KK spectrum we can reorganize it into superconformal multiplets. For the periodic spin structure the result was derived in \cite{Nilsson:1984bj,Liu:2016dau} and the resulting spectrum is given in equation \eqref{N6multiplets}. In this work we are more interested in the antiperiodic spin structure for which the fermionic spectrum is different and the full KK spectrum organizes into $\cN=2$ multiplets. In equation \eqref{N2multiplets} we presented the general result for any even KK level $n$. In Tables~\ref{tab:n2altmultipletapp} - \ref{tab:n8altmultipletapp} we illustrate this general result by explicitly listing the $\mathcal{N}=2$ superconformal multiplets, along with their $\SU(4)$ flavor symmetry representations, that arise from the first four even KK levels $n=2,4,6,8$.

\begin{table}[!htb]
\begin{minipage}[t]{0.45\linewidth}
		\centering
		\begin{tabular}{c|c}
			$\mathbf{1}$ &  $L\overline{L}[0]_4^{(0)}$ \Tstrut\\
			\hline
			$\mathbf{15}$ & $L\overline{L}[2]_3^{(0)}$ \Tstrut\\
			\hline
			$\mathbf{20}^\prime$ & $L\overline{L}[0]_3^{(0)}$\Tstrut\\
			\hline
			$\mathbf{35}$  & $L\overline{B}_1 [0]_2^{(2)}$ \Tstrut\\  
			\hline
			$\overline{\mathbf{35}}$ &  $B_1\overline{L} [0]_2^{(-2)}$ \Tstrut\\
			\hline
			$\mathbf{45}$ &  $L\overline{A}_1[1]_{\f52}^{(1)}$\Tstrut\\
			\hline
			$\overline{\mathbf{45}}$ &  $A_1\overline{L}[1]_{\f52}^{(-1)}$ \Tstrut\\
			\hline
			$\mathbf{84}$ & $L\overline{L}[0]_2^{(0)}$ \Tstrut
		\end{tabular}
		\caption{The $\cN=2$ multiplets from level $n=2$.}
		\label{tab:n2altmultipletapp}
\end{minipage}
\hspace{0.5cm}
\begin{minipage}[t]{0.45\linewidth}
		\centering
		\begin{tabular}{c|c|c|c}
			$\mathbf{15}$ & $L\overline{L}[0]_5^{(0)}$ & $\mathbf{84}$ & $L\overline{L}[2]_4^{(0)}$ \Tstrut\Bstrut\\
			\hline
			$\mathbf{175}$ & $L\overline{L}[0]_4^{(0)}$ & $\mathbf{300}^\prime$ & $L\overline{L}[0]_3^{(0)}$ \Tstrut\Bstrut\\
			\hline
			$\mathbf{35}$ & $L\overline{A}_1[2]_4^{(2)}$ & $\overline{\mathbf{35}}$ & $A_1\overline{L}[2]_4^{(-2)}$ \Tstrut\Bstrut\\
			\hline
			$\mathbf{45}$ & $L\overline{L}[1]_{\f92}^{(1)}$ & $\overline{\mathbf{45}}$ & $L\overline{L}[1]_{\f92}^{(-1)}$ \Tstrut\Bstrut\\
			\hline
			$\mathbf{189}$ & $L\overline{A}_2[0]_3^{(2)}$ & $\overline{\mathbf{189}}$ & $A_2\overline{L}[0]_3^{(-2)}$ \Tstrut\Bstrut\\
			\hline
			$\mathbf{256}$ & $L\overline{L}[1]_{\f72}^{(1)}$ & $\overline{\mathbf{256}}$ & $L\overline{L}[1]_{\f72}^{(-1)}$ \Tstrut\Bstrut\\
		\end{tabular}
		\caption{The $\cN=2$ multiplets from level $n=4$.}
		\label{tab:n4altmultipletapp}
\end{minipage}
\end{table}
\begin{table}[!htb]
	\begin{minipage}[t]{0.45\linewidth}
		\centering
		\begin{tabular}{c|c|c|c}
			$\mathbf{35}$ & $L\overline{L}[0]_6^{(2)}$ & $\overline{\mathbf{35}}$ & $L\overline{L}[0]_6^{(-2)}$ \Tstrut\Bstrut\\
			\hline
			$\mathbf{165}$ & $L\overline{B}_1[0]_4^{(4)}$ & $\overline{\mathbf{165}}^\prime$ & $B_1\overline{L}[0]_4^{(-4)}$ \Tstrut\Bstrut\\
			\hline
			$\mathbf{189}$ & $L\overline{L}[2]_5^{(2)}$ & $\overline{\mathbf{189}}$ & $L\overline{L}[2]_5^{(-2)}$ \Tstrut\Bstrut\\
			\hline
			$\mathbf{256}$ & $L\overline{L}[1]_{\f92}^{(1)}$ & $\overline{\mathbf{256}}$ & $L\overline{L}[1]_{\f92}^{(-1)}$ \Tstrut\Bstrut\\
			\hline
			$\mathbf{84}$ & $L\overline{L}[0]_6^{(0)}$ & $\mathbf{300}^\prime$ & $L\overline{L}[2]_5^{(0)}$ \Tstrut\Bstrut\\
			\hline
			$\mathbf{315}$ & $L\overline{A}_1[1]_{\f92}^{(3)}$ & $\overline{\mathbf{315}}$ & $A_1\overline{L}[1]_{\f92}^{(-3)}$ \Tstrut\Bstrut\\
			\hline
			$\mathbf{360}^\prime$ & $L\overline{L}[0]_{5}^{(2)}$ & $\overline{\mathbf{360}^\prime}$ & $L\overline{L}[0]_{5}^{(-2)}$ \Tstrut\Bstrut\\
			\hline
			$\mathbf{616}$ & $L\overline{L}[0]_{4}^{(2)}$ & $\overline{\mathbf{616}}$ & $L\overline{L}[0]_{4}^{(-2)}$ \Tstrut\Bstrut\\
			\hline
			$\mathbf{729}$ & $L\overline{L}[0]_{5}^{(0)}$ & $\mathbf{825}$ & $L\overline{L}[0]_{4}^{(0)}$ \Tstrut\Bstrut\\
			\hline
			$\mathbf{875}$ & $L\overline{L}[1]_{\f92}^{(1)}$ & $\overline{\mathbf{875}}$ & $L\overline{L}[1]_{\f92}^{(-1)}$ \Tstrut\Bstrut\\
		\end{tabular}
		\caption{The $\cN=2$ multiplets from level $n=6$.}
		\label{tab:n6altmultipletapp}
	\end{minipage}
	\hspace{0.5cm}
	\begin{minipage}[t]{0.45\linewidth}
		\centering
		\begin{tabular}{c|c|c|c}
			$\mathbf{165}$ & $L\overline{A}_1[2]_6^{(4)}$ & $\overline{\mathbf{165}}$ & $A_1\overline{L}[2]_6^{(-4)}$ \Tstrut\Bstrut\\
			\hline
			$\mathbf{189}$ & $L\overline{L}[0]_7^{(2)}$ & $\overline{\mathbf{189}}$ & $L\overline{L}[0]_7^{(-2)}$ \Tstrut\Bstrut\\
			\hline
			$\mathbf{315}$ & $L\overline{L}[1]_{\f{13}2}^{(3)}$ & $\overline{\mathbf{315}}$ & $L\overline{L}[1]_{\f{13}2}^{(-3)}$ \Tstrut\Bstrut\\
			\hline
			$\mathbf{616}$ & $L\overline{L}[2]_{6}^{(2)}$ & $\overline{\mathbf{616}}$ & $L\overline{L}[2]_{6}^{(-2)}$ \Tstrut\Bstrut\\
			\hline
			$\mathbf{715}$ & $L\overline{A}_2[0]_{5}^{(4)}$ & $\overline{\mathbf{715}}$ & $A_2\overline{L}[0]_{5}^{(-4)}$ \Tstrut\Bstrut\\
			\hline
			$\mathbf{875}$ & $L\overline{L}[1]_{\f{13}2}^{(1)}$ & $\overline{\mathbf{875}}$ & $L\overline{L}[1]_{\f{13}2}^{(-1)}$ \Tstrut\Bstrut\\
			\hline
			$\mathbf{1280}^{\prime\prime}$ & $L\overline{L}[1]_{\f{11}2}^{(3)}$ & $\overline{\mathbf{1280}^{\prime\prime}}$ & $L\overline{L}[1]_{\f{11}2}^{(-3)}$ \Tstrut\Bstrut\\
			\hline
			$\mathbf{1485}$ & $L\overline{L}[0]_{6}^{(2)}$ & $\overline{\mathbf{1485}}$ & $L\overline{L}[0]_{6}^{(-2)}$ \Tstrut\Bstrut\\
			\hline
			$\mathbf{1560}$ & $L\overline{L}[0]_{5}^{(2)}$ & $\overline{\mathbf{1560}}$ & $L\overline{L}[0]_{5}^{(-2)}$ \Tstrut\Bstrut\\
			\hline
			$\mathbf{2304}$ & $L\overline{L}[1]_{\f{11}2}^{(1)}$ & $\overline{\mathbf{2304}}$ & $L\overline{L}[1]_{\f{11}2}^{(-1)}$ \Tstrut\Bstrut\\
			\hline
			$\mathbf{300}^\prime$ & $L\overline{L}[0]_{7}^{(0)}$ & $\mathbf{825}$ & $L\overline{L}[2]_{6}^{(0)}$ \Tstrut\Bstrut\\
			\hline
			$\mathbf{1911}$ & $L\overline{L}[0]_{5}^{(0)}$ & $\mathbf{2156}$ & $L\overline{L}[0]_{6}^{(0)}$ \Tstrut\Bstrut\\
			\hline
		\end{tabular}
		\caption{The $\cN=2$ multiplets from level $n=8$.}
		\label{tab:n8altmultipletapp}
	\end{minipage}
\end{table}

\FloatBarrier

\subsection{Higher levels for the generalizations}

In Section~\ref{subsec:generalization1} and \ref{subsec:generalization2} we introduced a class of generalizations to $\bZ_{4p}$ orbifolds and formulated a rule for selecting the invariant spectrum for both spin structures for the orbifold defined by the action~\eqref{first}. In the main text we illustrated this rule by explicitly showing the resulting spectrum at KK level $n=0$ and showed that it reorganizes into $\cN=4$ or $\cN=2$ superconformal multiplets, depending on the chosen spin structure.

Here we give further evidence for this rule and give the explicit KK spectrum at level $n=2$ for the orbifold with action \eqref{first}. The branching rules of the relevant $\SO(8)$ representations under $\SU(2)_1\times \SU(2)_2\times \UU(1)_{\SU(4)}\times \UU(1)$ are rather lengthy so we first present the invariant spectrum for the bosons which is identical for both spin structures and next discuss the fermionic spectrum separately. In addition, we take $k\geq 12$ in which case we only select the neutral part of the spectrum at level $n$. For $k=8$ and $k=4$ there are a few additional representations appearing which are charged under the $\UU(1)$ flavor symmetry. At level $n=0$ we saw the same thing happening where for $k=4$ an additional pair of SUSY current multiplets appeared.

The relevant representations for level $n=2$ are given in Table~\ref{tab:multipletn}. The branching rules for the bosonic modes are given by:
\begin{align}
	(2,0,0,0)\rightarrow & \, \left(\mathbf{1},\mathbf{1}\right)_{0,0}\oplus\left(\mathbf{1},\mathbf{3}\right)_{0,0}\oplus\left(\mathbf{3},\mathbf{1}\right)_{0,0}\,,\\[2mm]
	(2,1,0,0)\rightarrow & \,2\left(\mathbf{1},\mathbf{1}\right)_{0,0}\oplus 4\left(\mathbf{1},\mathbf{3}\right)_{0,0}\oplus 4\left(\mathbf{3},\mathbf{1}\right)_{0,0}\nn\\
	& \oplus 3\left(\mathbf{3},\mathbf{3}\right)_{0,0} \oplus\left(\mathbf{1},\mathbf{5}\right)_{0,0} \oplus\left(\mathbf{5},\mathbf{1}\right)_{0,0}\\
	&\oplus \left(\mathbf{1},\mathbf{3}\right)_{0,\pm 1} \oplus\left(\mathbf{3},\mathbf{1}\right)_{0,\pm 1} \oplus\left(\mathbf{3},\mathbf{3}\right)_{0,\pm 1}\,,\nn\\[2mm]
	(1,0,1,1)\rightarrow & \,3\left(\mathbf{1},\mathbf{1}\right)_{0,0}\oplus \left(\mathbf{1},\mathbf{1}\right)_{0,\pm 1} \nn\\
	&\oplus 4\left(\mathbf{1},\mathbf{3}\right)_{0,0} \oplus 4\left(\mathbf{3},\mathbf{1}\right)_{0,0} \oplus 3\left(\mathbf{3},\mathbf{3}\right)_{0,0}\\
	&\oplus \left(\mathbf{1},\mathbf{3}\right)_{0,\pm 1} \oplus\left(\mathbf{3},\mathbf{1}\right)_{0,\pm 1}\,,\nn\\
	(0,1,0,0)\rightarrow& 2\left(\mathbf{1},\mathbf{1}\right)_{0,0} \oplus \left(\mathbf{1},\mathbf{3}\right)_{0,0} \oplus \left(\mathbf{3},\mathbf{1}\right)_{0,0} \,,\\[2mm]
	(4,0,0,0)\rightarrow &\, \left(\mathbf{1},\mathbf{1}\right)_{0,0} \oplus \left(\mathbf{3},\mathbf{3}\right)_{0,\pm 1}\nn \\
	&\oplus \left(\mathbf{1},\mathbf{3}\right)_{0,0} \oplus \left(\mathbf{3},\mathbf{1}\right)_{0,0} \oplus \left(\mathbf{3},\mathbf{3}\right)_{0,0}\\
	&\oplus \left(\mathbf{1},\mathbf{5}\right)_{0,0} \oplus\left(\mathbf{5},\mathbf{1}\right)_{0,0}\,,\nn\\[2mm]
	(0,2,0,0)\rightarrow & \,4\left(\mathbf{1},\mathbf{1}\right)_{0,0} \oplus 2\left(\mathbf{1},\mathbf{3}\right)_{0,0} \oplus 2\left(\mathbf{3},\mathbf{1}\right)_{0,0} \nn\\
	&\oplus 2\left(\mathbf{3},\mathbf{3}\right)_{0,0} \oplus \left(\mathbf{1},\mathbf{5}\right)_{0,0} \oplus \left(\mathbf{5},\mathbf{1}\right)_{0,0}\\
	&\oplus \left(\mathbf{1},\mathbf{1}\right)_{0,\pm 1} \oplus\left(\mathbf{3},\mathbf{3}\right)_{0,\pm 1}\,,\nn\\[2mm]
	(0,0,0,0)\rightarrow & \,\left(\mathbf{1},\mathbf{1}\right)_{0,0}\,,\\[2mm]
	(2,0,2,0)\rightarrow & \,2\left(\mathbf{1},\mathbf{1}\right)_{0,0} \oplus 3\left(\mathbf{1},\mathbf{3}\right)_{0,0} \oplus 3\left(\mathbf{3},\mathbf{1}\right)_{0,0} \nn\\
	&\oplus 6\left(\mathbf{3},\mathbf{3}\right)_{0,0} \oplus \left(\mathbf{1},\mathbf{5}\right)_{0,0} \oplus \left(\mathbf{5},\mathbf{1}\right)_{0,0}\\
	&\oplus \left(\mathbf{1},\mathbf{1}\right)_{0,\pm 1} \oplus\left(\mathbf{1},\mathbf{3}\right)_{0,\pm 1} \oplus\left(\mathbf{3},\mathbf{1}\right)_{0,\pm 1}\nn\\
	&\oplus \left(\mathbf{3},\mathbf{3}\right)_{0,\pm 1} \oplus\left(\mathbf{1},\mathbf{5}\right)_{0,\pm 1} \oplus\left(\mathbf{5},\mathbf{1}\right)_{0,\pm 1}\,,\nn\\[2mm]
	(0,0,0,2)\rightarrow & \,2\left(\mathbf{1},\mathbf{1}\right)_{0,0} \oplus \left(\mathbf{1},\mathbf{1}\right)_{0,\pm 1} \oplus \left(\mathbf{3},\mathbf{3}\right)_{0,0}\,.
\end{align}
Moving on to the fermionic part of the spectrum, we find that the modes invariant under the orbifold action for the periodic spin structure are given as follows:
\begin{align}
		(2,0,0,1)\rightarrow & \, 4\left(\mathbf{2},\mathbf{2}\right)_{0,0} \oplus \left(\mathbf{2},\mathbf{2}\right)_{0,\pm1} \oplus \left(\mathbf{2},\mathbf{4}\right)_{0,0}\oplus \left(\mathbf{4},\mathbf{2}\right)_{0,0}\,,\\
		(1,0,1,0)\rightarrow&\, 3\left(\mathbf{2},\mathbf{2}\right)_{0,0}\,,\\[2mm]
		(3,0,1,0)\rightarrow &\, 4\left(\mathbf{2},\mathbf{2}\right)_{0,0} \oplus 3\left(\mathbf{2},\mathbf{4}\right)_{0,0} \oplus 3\left(\mathbf{4},\mathbf{2}\right)_{0,0} \\
		& \oplus \left(\mathbf{2},\mathbf{2}\right)_{0,\pm1} \oplus \left(\mathbf{2},\mathbf{4}\right)_{0,\pm 1} \oplus \left(\mathbf{4},\mathbf{2}\right)_{0,\pm 1}\,,\nn\\[2mm]
		(1,1,1,0)\rightarrow &\, 9\left(\mathbf{2},\mathbf{2}\right)_{0,0} \oplus 4\left(\mathbf{2},\mathbf{4}\right)_{0, 0} \oplus 4\left(\mathbf{4},\mathbf{2}\right)_{0,0} \\
		&\oplus 2\left(\mathbf{2},\mathbf{2}\right)_{0,\pm 1} \oplus \left(\mathbf{2},\mathbf{4}\right)_{0,\pm 1 } \oplus \left(\mathbf{4},\mathbf{2}\right)_{0,\pm 1}\,,\nn\\[2mm]
		(0,1,0,1)\rightarrow & \,\left(\mathbf{2},\mathbf{2}\right)_{0,\pm 1} \oplus 4\left(\mathbf{2},\mathbf{2}\right)_{0,0} \oplus \left(\mathbf{2},\mathbf{4}\right)_{0,0} \oplus \left(\mathbf{4},\mathbf{2}\right)_{0,0}\,,\nn\\[2mm]
		(0,0,0,1)\rightarrow &\, \left(\mathbf{2},\mathbf{2}\right)_{0,0}\,.
\end{align}
For the antiperiodic spin structure on the other hand we find the following invariant modes:
\begin{align}
		(2,0,0,1)\rightarrow & \,\left(\mathbf{1},\mathbf{1}\right)_{0,\pm\f12} \oplus 2\left(\mathbf{1},\mathbf{3}\right)_{0,\pm\f12} \oplus 2\left(\mathbf{3},\mathbf{1}\right)_{0,\pm\f12} \oplus \left(\mathbf{3},\mathbf{3}\right)_{0,\pm\f12}\,,\\[2mm]
		(1,0,1,0)\rightarrow &\,\left(\mathbf{1},\mathbf{1}\right)_{0,\pm \f12} \oplus \left(\mathbf{1},\mathbf{3}\right)_{0,\pm\f12} \oplus \left(\mathbf{3},\mathbf{1}\right)_{0,\pm\f12}\,,\\[2mm]
		(3,0,1,0)\rightarrow &\, 3\left(\mathbf{3},\mathbf{3}\right)_{0,\pm\f12} \oplus 2\left(\mathbf{1},\mathbf{3}\right)_{0,\pm\f12} \oplus 2\left(\mathbf{3},\mathbf{1}\right)_{0,\pm\f12}\\
		& \oplus \left(\mathbf{1},\mathbf{1}\right)_{0,\pm\f12}\oplus \left(\mathbf{1},\mathbf{5}\right)_{0,\pm\f12} \oplus \left(\mathbf{5},\mathbf{1}\right)_{0,\pm\f12}\,,\nn\\[2mm]
		(1,1,1,0)\rightarrow & 3\left(\mathbf{1},\mathbf{1}\right)_{0, \pm \f12} \oplus 4\left(\mathbf{1},\mathbf{3}\right)_{0,\pm \f12} \oplus 4\left(\mathbf{3},\mathbf{1}\right)_{0,\pm \f12}\oplus 4\left(\mathbf{3},\mathbf{3}\right)_{0,\pm \f12}\\
		& \oplus \left(\mathbf{1},\mathbf{5}\right)_{0,\pm \f12}\oplus \left(\mathbf{5},\mathbf{1}\right)_{0,\pm \f12}\nn\,,\\[2mm]
		(0,1,0,1)\rightarrow &\, 3\left(\mathbf{1},\mathbf{1}\right)_{0,\pm\f12} \oplus \left(\mathbf{1},\mathbf{3}\right)_{0,\pm\f12} \oplus \left(\mathbf{3},\mathbf{1}\right)_{0,\pm\f12} \oplus \left(\mathbf{3},\mathbf{3}\right)_{0,\pm\f12}\,,\\[2mm]
		(0,0,0,1)\rightarrow &\, \left(\mathbf{1},\mathbf{1}\right)_{0,\pm\f12}\,.
\end{align}
We can again organize these modes in superconformal multiplets and find that, as expected, for the periodic spin structure they organize into $\cN=4$ multiplets, with all components having the same $\UU(1)$ charges, while for the antiperiodic spin structure they organize into $\cN=2$ multiplets. In Table~\ref{tab:n2genmult} we give the resulting $\cN=4$ multiplets for each combination of $\UU(1)^2$ flavor charges.
\begin{table}[!htb]
	\centering
	\begin{tabular}{c|c}
		$\UU(1)_{\SU(4)}\times \UU(1)$ & $\cN=4$ supermultiplets\\
		\hline
		$\left(0,0\right)$ & $B_1[0]_2^{\left(\mathbf{3},\mathbf{3}\right)}\,,\quad B_1[0]_2^{\left(\mathbf{1},\mathbf{5}\right)}\,,\quad B_1[0]_2^{\left(\mathbf{5},\mathbf{1}\right)}\,,\quad A_2[0]_2^{\left(\mathbf{0},\mathbf{2}\right)}\,,\quad A_2[0]_2^{\left(\mathbf{2},\mathbf{0}\right)}\,,\quad L[0]_2^{\left(\mathbf{1},\mathbf{1}\right)} \Tstrut$\\
		\hline
		$\left(0,1\right)$ & $B_1[0]_2^{\left(\mathbf{3},\mathbf{3}\right)}\Tstrut$\\
		\hline
		$\left(0,-1\right)$ & $B_1[0]_2^{\left(\mathbf{3},\mathbf{3}\right)}\Tstrut$
		\end{tabular}
	\caption{The $\cN=4$ multiplets from the KK spectrum at level $n=2$ for the periodic spin structure. The $\UU(1)_{\SU(4)}\times \UU(1)$ charges are presented in the left column while the $\SU(2)_1 \times \SU(2)_2$ representations are indicated on each multiplet.}
	\label{tab:n2genmult}
\end{table}
Similarly, for the antiperiodic spin structure the bosonic and fermionic modes organize into $\cN=2$ multiplets where all components of the superconformal multiplet have the same $\SU(2)_1\times \SU(2)_2\times\UU(1)_{\SU(4)}$ charges. In Table~\ref{tab:n2genaltmult} we give the resulting $\cN=2$ multiplets for each $\SU(2)_1\times \SU(2)_2\times \UU(1)_{\SU(4)}$ representation.
\begin{table}[!htb]
	\centering
	\begin{tabular}{c|c}
		Flavor representation & $\cN=2$ supermultiplets\\
		\hline
		$\left(\mathbf{1},\mathbf{1}\right)_0$ & $L\overline{L}[2]_3^{(0)}\,,\quad L\overline{L}[0]_2^{(0)}\,,\quad L\overline{L}[0]_3^{(0)}\,,\quad L\overline{L}[0]_4^{(0)}\Tstrut$\\
		\hline
		$\left(\mathbf{1},\mathbf{3}\right)_0$ & $L\overline{L}[2]_3^{(0)}\,,\quad L\overline{L}[0]_2^{(0)}\,,\quad A_1\overline{L}[1]_{\f52}^{(\f15)}\,,\quad L\overline{A}_1[1]_{\f52}^{(-\f12)}\Tstrut$\\
		\hline
		$\left(\mathbf{3},\mathbf{1}\right)_0$ & $L\overline{L}[2]_3^{(0)}\,,\quad L\overline{L}[0]_2^{(0)}\,,\quad A_1\overline{L}[1]_{\f52}^{(\f15)}\,,\quad L\overline{A}_1[1]_{\f52}^{(-\f12)}\Tstrut$\\
		\hline
		\multirow{2}{*}{$\left(\mathbf{3},\mathbf{3}\right)_0$} & $L\overline{L}[0]_2^{(0)}\,,\quad L\overline{L}[0]_3^{(0)}\,,\quad B_1\overline{L}[0]_2^{(1)}\,,\Tstrut$\\
		&  $L\overline{B}_1[0]_2^{(-1)}\,,\quad A_1\overline{L}[1]_{\f52}^{(\f15)}\,,\quad L\overline{A}_1[1]_{\f52}^{(-\f12)}\Tstrut$\\
		\hline
		$\left(\mathbf{1},\mathbf{5}\right)_0$ & $L\overline{L}[0]_2^{(0)}\Tstrut$\\
		\hline
		$\left(\mathbf{5},\mathbf{1}\right)_0$ & $L\overline{L}[0]_2^{(0)}\Tstrut$
	\end{tabular}
\caption{The $\cN=2$ multiplets from the KK spectrum at level $n=2$ for the antiperiodic spin structure. The $\SU(2)_1\times \SU(2)_2 \times \UU(1)_{\SU(4)}$ flavor symmetry representations are indicated in the left column.}
\label{tab:n2genaltmult}
\end{table}

\clearpage
\FloatBarrier
\bibliography{allpapers}
\bibliographystyle{JHEP}
	
\end{document}